\documentclass{iopjournal}
\usepackage[T1]{fontenc}
\usepackage{lmodern}
\usepackage{amsmath,amssymb,amsthm}
\usepackage{booktabs}
\usepackage{cite}
\newcommand{\ee}{\mathrm{e}}
\newcommand{\htop}{h_{\rm top}}
\newcommand{\wc}{w_c}
\newtheorem{lemma}{Lemma}

\makeatletter
\renewcommand{\articletype}[1]{{\vspace*{-8mm}\noindent\Large\sf J. Phys. A: Math. Theor.}\par
  \vspace*{8mm}\noindent{\scriptsize\sf\bfseries\MakeUppercase{#1}}}
\makeatother
\usepackage{microtype}
\usepackage{ragged2e}
\AtBeginDocument{\justifying}
\makeatletter
\renewenvironment{abstract}{%
      \vspace{16pt plus3pt minus3pt}
      {\color{gray}\hrule} \ \\
      \noindent \fontsize{11}{12}\selectfont {\bfseries Abstract}\\
      \rm\ignorespaces \justifying}{\vspace{3mm} {\color{gray}\hrule}}
\renewcommand{\@makecaption}[2]{\vskip\abovecaptionskip
\sbox\@tempboxa{\fontsize{8}{10}\selectfont {\bfseries #1.} #2}%
\ifdim \wd\@tempboxa >\hsize
\justifying\noindent \fontsize{8}{10}\selectfont {\bfseries #1.} #2\par
\else
\global \@minipagefalse
\hb@xt@\hsize{\hfil\box\@tempboxa\hfil}%
\fi
\vskip\belowcaptionskip}
\makeatother

\begin{document}

\articletype{Paper}

\title{The Thue--Morse route to chaos in the open quantum baker map}

\author{Leonardo Ermann$^{1,2}$\orcid{0000-0003-0964-3929}}

\affil{$^1$DFTIFSC, GIyA, Comisi\'on Nacional de Energ\'ia At\'omica (CNEA) and CONICET, Buenos Aires, Argentina}

\affil{$^2$ECyT, Universidad Nacional de General San Mart\'in (UNSAM), B1653HIM San Mart\'in, Argentina}

\email{leonardoermann@cnea.gob.ar}


\begin{abstract}
We study the resonances of the quantum baker map opened by a window of width $w$ centred at $q_c$. For $q_c=1/2$ the
survivor set of the classical doubling map with a hole goes from zero to positive topological entropy along the
Thue--Morse route, and we find numerically, for $N$ up to $2^{25}$, that the resonance spectrum follows this route step by
step, also where the entropy is zero. Below the threshold $\wc=4a^*-1$, with $a^*$ the Thue--Morse constant, the surviving
cycles have periods $2^\nu$ and their thresholds accumulate doubly exponentially. Each cycle gives, as isolated orbits do, a multiplet of
resonances close to modulus $2^{-1/2}$, with phases set by its action. Since a cycle is resolved only if its period is
shorter than about $\log_2N$, twice the Ehrenfest time, about $\log_2\log_2N$ steps are resolved, four at our largest sizes. Near $\wc$ the 16-cycle perturbs the 8-multiplet
through a return path that classically opens only at $\wc$. Cutting a few sites of this path removes the
perturbation. At $\wc$ the leading resonance jumps within one or two lattice sites, since a window one site wider than the
threshold already has an entropy of order $1/\log_2N$. One site or more above $\wc$ it lies between $0.93$ and $1.00$ times the value
set by the topological pressure, with no free parameter. There the Walsh eigenvalue count grows
with the fractal Weyl exponent, and the Fourier count with an exponent a few hundredths smaller at the sizes we
reach.
\end{abstract}

\section{Introduction}\label{sec:intro}

The resonance spectrum of open quantum maps with chaotic classical dynamics is organised by the trapped set,
the set of orbits that never escape. When the trapped set is a hyperbolic repeller of dimension $D$, the number of
resonances above a fixed modulus is expected to grow with the Hilbert space dimension $N$ as $N^{D/2}$, the so-called
fractal Weyl law \cite{Sjostrand1990,LuSridharZworski2003}. An upper bound of this order is proven for open quantum maps
\cite{NSZ2014}, and the leading resonance is bounded by the topological pressure of the repeller
\cite{NonnenmacherZworski2009,Nonnenmacher2011}. For the open quantum baker map the fractal Weyl law was confirmed
numerically, and proven for a Walsh quantisation with openings made of digit strips \cite{NZ2005,NZ2007}. The
distribution of resonances depends strongly on the position of the opening \cite{Pedrosa2009}, as do the classical
escape rates of leaking systems \cite{BunimovichYurchenko2011,Altmann2013}; see \cite{SchomerusTworzydlo2004} for the
open kicked rotator, \cite{Ermann2026PRE} for the spectral statistics of a partially open baker map and
\cite{Novaes2013} for a review. All these results concern repellers of positive entropy. The
opposite limit is much less explored. We recently showed that the classical pruning tiles the $(q_c,w)$ plane into
regions where a single periodic orbit survives, and that this orbit gives a multiplet of resonances of modulus close
to $2^{-1/2}$, the value set by the topological pressure, with phases quantised by its action \cite{Ermann2026PRL}.
Where two isolated orbits coexist, as at $q_c=1/2$ for $3/5\le w\lesssim0.647$, the spectrum contains both multiplets.
It is not known what happens along a whole cascade of orbits, nor how the spectrum changes as the window grows until
the entropy becomes positive.

The doubling map with a hole is the minimal model for this question. Let $\sigma x=2x\bmod1$ be the doubling map, the
shift on binary expansions, and let $H$ be an open interval. The survivor set
$\mathcal K(H)=\{x:\,\sigma^t x\notin H\ \forall t\ge0\}$ (written $\mathcal J$ in \cite{GlendinningSidorov2015}) has
been characterised in detail in the mathematical literature, which we review in section~\ref{sec:classical}. The map
is chaotic for every hole; what changes with the hole is the survivor set, and by a route to chaos we mean here the
passage of $\mathcal K$ from zero to positive topological entropy as the hole shrinks. For
symmetric holes it has positive Hausdorff dimension if and only if the hole is shorter than a critical value fixed by
the Thue--Morse sequence \cite{GlendinningSparrow1993,ZyczkowskiBollt1999,Moreira2001,Nilsson2010,GlendinningSidorov2015}.
For general holes Labarca and Moreira characterised the boundary between zero and positive entropy in the plane of
the hole endpoints, and Glendinning and Sidorov gave it explicitly \cite{LabarcaMoreira2006,GlendinningSidorov2015}.
All these results concern the survivor set only; to our knowledge, the quantum counterpart has not been studied. In
the quantum baker map the Thue--Morse sequence has appeared in another role: for $N=2^n$ some eigenstates of the
closed map are dominated by it in the Walsh--Hadamard basis and are scarred by the period-2 orbit \cite{ML2005,LM2006}.

In this work we combine the classical theory with numerical diagonalisation of the open baker map (dense, and
Arnoldi up to $N=2^{25}$), and we find
that the resonance spectrum follows the classical route to chaos step by step, also where the entropy is zero. This
was not guaranteed: near $\wc$ the thresholds of successive cycles are only a few lattice sites apart at the sizes we
can reach, and diffraction at the window edges could have merged the multiplets or smeared the transition. On
the classical side we collect what the route looks like. For the symmetric window the transition to positive entropy
is preceded by an infinite cascade of cycles of period $2^\nu$, the Thue--Morse cycles. Their thresholds are known in
closed form \cite{ZyczkowskiBollt1999,GlendinningSidorov2015,ACS2009}, and \.Zyczkowski and Bollt noted that they
accumulate doubly exponentially, unlike the geometric accumulation of the Feigenbaum cascade
\cite{ZyczkowskiBollt1999}. We give the prefactor and the first correction of this law, and we check the thresholds by
an exhaustive search of all cycles of period up to $32$. Above $\wc$ the entropy is a devil's staircase whose plateaus
accumulate at $\wc$. From the results of Nilsson \cite{Nilsson2010} and of Alcaraz Barrera, Baker and Kong
\cite{AlcarazBakerKong} we evaluate five plateaus near $\wc$ in closed form and certify them in exact integer
arithmetic. Since the entropy cannot decrease when the window grows, these plateaus bound it on both sides over six
decades of $w-\wc$, $1.24\le\htop\log_2[1/(w-\wc)]\le2.65$, so that a window one lattice site above $\wc$, at
$w-\wc\simeq1/N$, has an entropy of order $1/\log_2N$. We claim no new classical theorem: these results rest on the theorems cited above, which we apply to our
opening through an elementary reduction (lemma~\ref{lem:reduction}).

The results for the Fourier quantisation are numerical, with the accuracy stated in each case. Throughout, the Walsh
quantisation \cite{NZ2005,NZ2007} serves as a reference without diffraction: its open spectrum is fixed by the
lattice graph of the window (section~\ref{sec:model}), and we attribute to diffraction the differences between the
two quantisations, keeping in mind that the Walsh baker has arithmetic features of its own. Below $\wc$ each surviving
cycle produces a multiplet of resonances whose phases are set by its action, so the number of resolved resonances
steps up at the threshold of each cycle (figure~\ref{fig:route}). A cycle is resolved only when its period is shorter
than about $n=\log_2N$, twice the Ehrenfest time \cite{BermanZaslavsky1978}, and since the periods double at each step a Hilbert
space of dimension $N$ resolves only about $\log_2\log_2N$ steps of the cascade. Close to $\wc$ the 16-cycle perturbs the 8-multiplet through a return path
that classically opens only at $\wc$, and removing a few sites of this path removes the perturbation; we attribute the
coupling to diffraction, and this finite-$N$ effect is a quantum precursor of the transition. At $\wc$ the leading
resonance jumps within one or two lattice sites, by an amount that decreases slowly with $N$. One site or more above
$\wc$ it lies between the value $\ee^{\htop/2}/\sqrt2$ given by the classical decay of the survival probability and
the value $\ee^{P(1/2)}=\ee^{\htop}/\sqrt2$ set by the topological pressure, at $0.93$ to $1.00$ times the latter, with
no free parameter. The jump itself lies between its classical and its pressure values, and at several sizes it is
close to the classical one. It falls short of the pressure value because above the threshold $|\lambda_0|$ stays
slightly below $\ee^{P(1/2)}$, and a few sites below the threshold it already exceeds $2^{-1/2}$.
At the end of the route, where the entropy is positive, the Walsh eigenvalue count follows the fractal Weyl law, and
the Fourier count grows with a slightly smaller exponent at the sizes we reach. Below $\wc$ the fractal Weyl law
predicts a number of resonances independent of $N$; how many resonances the zero-entropy windows actually carry is the
subject of a separate paper \cite{ErmannP18}.

The paper is organised as follows. Section~\ref{sec:model} defines the model and the opening and reduces holes that
contain the fixed point to the form used in the classical literature. Section~\ref{sec:classical} collects the
classical results: the boundary of the zero-entropy region, the Thue--Morse cascade and the entropy above $\wc$.
Section~\ref{sec:quantum} presents the quantum results along the route, namely the multiplets below $\wc$, their
hybridisation near $\wc$, the jump of $|\lambda_0|$ at $\wc$, the comparison with the pressure and the resonance
count above $\wc$. Section~\ref{sec:discussion} contains the discussion. Appendix~\ref{app:lemma} contains the
reduction lemma, appendix~\ref{app:cert} the origin and the exact certificates of the entropy plateaus,
appendix~\ref{app:D} the operator behind an $N$-independent eigenvalue, and appendix~\ref{app:routes} other routes to
the boundary.

\section{Model}\label{sec:model}

Let $N$ be even ($N=2^n$ unless stated otherwise) and let $G_N$ be the discrete Fourier transform of dimension $N$ with half-integer shifts
(antiperiodic boundary conditions) \cite{Saraceno1990}. The quantum baker map \cite{BalazsVoros1989,Saraceno1990} is
$U=G_N^\dagger\,\mathrm{diag}(G_{N/2},G_{N/2})$. The opened map is $\mathcal U=U\,\Pi_W$, where $\Pi_W$ projects on
a window $W$ of $M=\mathrm{round}(wN)$ consecutive sites centred at $q_c$, the sites
$j_c-\lfloor M/2\rfloor,\dots,j_c-\lfloor M/2\rfloor+M-1$ with $j_c=\mathrm{round}(q_cN)$. The nonzero eigenvalues of
$\mathcal U$ are the eigenvalues of the $M\times M$ block $\Pi_WU\Pi_W$. In this paper $q_c=1/2$ unless stated
otherwise. For an eigenvalue $\lambda$ of the block we write $\psi^R$ for its right eigenvector.

The Walsh quantisation \cite{NZ2005,NZ2007}, defined for $N=2^n$, is the matrix whose only nonzero elements are
\begin{equation}
\langle 2j+\epsilon'\bmod N|\,U_{\rm W}\,|j\rangle=\frac{(-1)^{\epsilon\epsilon'}}{\sqrt2},\qquad
\epsilon=\lfloor2j/N\rfloor,\quad\epsilon'=0,1,
\label{eq:walsh}
\end{equation}
with $\epsilon$ the first binary digit of $j$. It moves amplitude only along the edges $j\to2j+\epsilon'\bmod N$, so
that its open spectrum is fixed by the lattice graph of the window, the restriction of this directed graph to the
sites of $W$. Its nonzero eigenvalues live on the non-trivial strongly connected components.

The parity $\mathcal R:\,|j\rangle\to|N-1-j\rangle$, i.e.\ $q\to1-q$, commutes with $U$. A window centred at
$q_c=1/2$ is invariant under $\mathcal R$ when $M$ is even, and then $\mathcal R$ commutes with $\mathcal U$ and the
spectrum splits into two independent parity sectors. For odd $M$ the lattice window is displaced by half a site and
the symmetry is broken at that scale: $[\mathcal R,\mathcal U]=U(\Pi_{W'}-\Pi_W)\mathcal R$, with $W'$ the window
shifted by one site, whose Frobenius norm is $\sqrt2$. For $q_c\ne1/2$ (and $q_c\ne0$) the open map has no parity.
We use both even and odd $M$. The half-site shift between them does not change the multiplets
(section~\ref{sec:mult}), but near $\wc$ it changes the entropy of the lattice window (section~\ref{sec:jump}).

Equation~\eqref{eq:walsh} is the Walsh quantisation of \cite{NZ2005,NZ2007}, built on the periodic $2\times2$
transform, the Hadamard matrix, while $G_N$ is antiperiodic. With the antiperiodic $G_2$ in its place the Walsh baker
commutes with $\mathcal R$, as $U$ does; equation~\eqref{eq:walsh} does not, because $\mathcal R$ maps a cycle onto
its complement and the Hadamard weights multiply the weight of the cycle by $(-1)^T$. We keep the periodic version, in which the fractal Weyl law was proven for openings made of
digit strips, because the choice only changes the phases of the weights. In the zero-entropy windows of this paper
every strongly connected component is a single cycle, whose eigenvalues are the $T$-th roots of the product of its
weights, so the change rotates each multiplet by an angle independent of $N$; in the windows of
section~\ref{sec:weyl} the counts with $|\lambda|>0.2$ agree within $2\%$ for $n\le16$ and the exponents within
$2\times10^{-3}$.

Where all eigenvalues are needed we diagonalise the block densely (complete spectra up to $N=2^{17}$). For the
leading eigenvalues at larger $N$ we compute the $6$ to $40$ eigenvalues of largest modulus with the implicitly restarted Arnoldi method on the $M\times M$ block, applying $U$ with fast Fourier
transforms, so that the matrix is never stored. The two methods agree to $10^{-12}$ for $|\lambda_0|$ at $N=1024$ and
$2048$, and the residuals of the Arnoldi eigenvalues are below $10^{-10}$ in every case.

Classically the map is the baker map, whose dynamics along the position coordinate $q\equiv x\in[0,1)$ is the
doubling map $\sigma x=2x\bmod 1$; the survivor set is $\mathcal K=\{x:\,\sigma^t x\in W\ \forall t\ge0\}$. The hole is
$H=[0,1)\setminus W$, an interval of length $1-w$ centred at $q_c+1/2$. The results of
\cite{GlendinningSidorov2015} assume a hole $(a,b)$ with $0<a<b<1$. Holes that contain the fixed point $0$ reduce to
this case through one preimage (lemma~\ref{lem:reduction} in appendix~\ref{app:lemma}). Labarca and Moreira \cite{LabarcaMoreira2006} and Glendinning and Sparrow
\cite{GlendinningSparrow1993} work with holes that contain $0$; we need the reduction to use the explicit results of
\cite{GlendinningSidorov2015}.

The lemma covers windows with $w\ge1/2$, which is all we need. For $q_c=1/2$ the reduced hole is $H'=(a,1-a)$ with $1-2a=h=(1-w)/2$, and $x\in\mathcal K$ iff $\|2^tx\|\ge h$ for all $t$,
where $\|\cdot\|$ is the distance to the nearest integer.

\section{Classical results}\label{sec:classical}

The quantum part of the paper uses three classical facts: the threshold $\wc$, equation~\eqref{eq:wc}, the doubly
exponential accumulation of the cycle thresholds, equation~\eqref{eq:dexp}, and the bound on the entropy above $\wc$,
equation~\eqref{eq:hstair}. A reader interested mainly in the quantum results can take these three equations and go
to section~\ref{sec:quantum}. They rest on a large literature on the doubling map with a hole. Urba\'nski related the
Hausdorff dimension of the set of points whose orbits avoid an interval $(0,\epsilon)$ to the topological entropy and
studied its dependence on $\epsilon$ \cite{Urbanski1986}. For the symmetric hole Nilsson described the dimension as
a function of the hole length \cite{Nilsson2009}, and later its fine structure \cite{Nilsson2010}, including the
intervals on which it is constant, which we use in section~\ref{sec:entropy}. For this hole, cycles appear as the hole shrinks in the
reverse of Sharkovski\u{\i}'s order, a consequence of the work of Allouche, Clarke and Sidorov on unique
$\beta$-expansions \cite{ACS2009,GlendinningSidorov2015}. In that problem the Thue--Morse sequence fixes the smallest
base in which $1$ has a unique expansion \cite{KomornikLoreti1998}, and at that base the set of numbers with a unique
expansion becomes uncountable \cite{GlendinningSidorov2001}; see also \cite{HareSidorov2014,Sidorov2014}. Recent work
extends these results to the $\beta$-transformation with a hole \cite{KongPu2026} and to the negative doubling map
\cite{DingSteinerSun2026}.

\subsection{Boundary of the zero-entropy region}\label{sec:boundary}

\begin{figure}
\centering
\includegraphics[width=\textwidth]{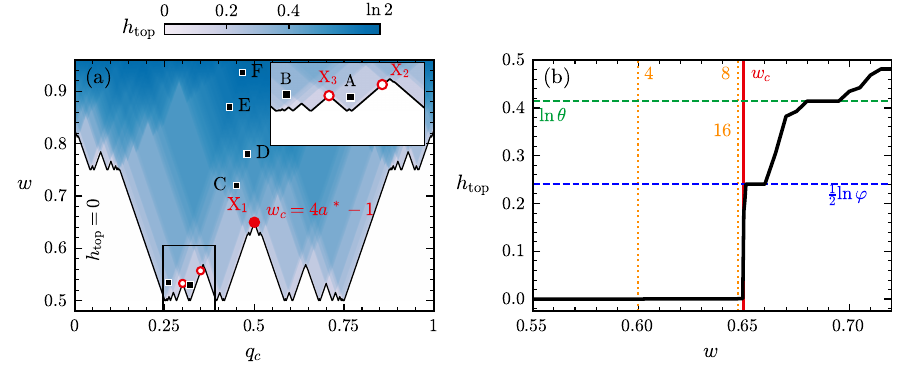}
\caption{Classical results. (a) Topological entropy of the doubling map with a window of width $w$ centred at $q_c$
(a hole of length $1-w$ centred at $q_c+1/2$), from the outer SFT bound at $m=18$ on a grid of $2001\times2001$
windows with $0\le q_c\le1/2$, mirrored by $q_c\to1-q_c$ (colour scale above the panel; white: $\htop=0$). Near the boundary the colour is an upper bound (see the
text). Full line: the boundary $\wc(q_c)$ of the zero-entropy region. Symbols mark
the windows used in this paper: dot and circles, the thresholds X$_1$, X$_2$, X$_3$ crossed by the jump scans at
$q_c=1/2,\,0.35,\,0.30$ (section~\ref{sec:jump}; X$_1$ is $\wc=4a^*-1$, equation~\eqref{eq:wc}); squares, the windows
A--F of section~\ref{sec:weyl}. X$_1$ and X$_3$ lie at tips of the boundary, X$_2$ on a flank of slope $2$, where only the
left edge of the window, at $1/14$, is critical. The inset magnifies the framed region. (b) Topological entropy versus $w$ at $q_c=1/2$ (outer SFT bound at $m=18$, estimated by power iteration). Dotted lines: appearance thresholds
$w_\nu$ of the Thue--Morse cycles of periods $4,8,16$ (the period-16 line lies $1.1\times10^{-5}$ below $\wc$ and is not
resolved at this scale); full line: $\wc$; dashed lines: the plateaus $\tfrac12\ln\varphi$, equation~\eqref{eq:plateau},
and $\ln\theta$, equation~\eqref{eq:plateau3}.}
\label{fig:cl}
\end{figure}

Labarca and Moreira showed that $\dim_H \mathcal K(a,b)>0$ if and only if $b<\chi(a)$ (\cite{LabarcaMoreira2006}, Theorem~2;
\cite{GlendinningSidorov2015}, Theorem~2.3). Glendinning and Sidorov gave $\chi$ explicitly
(\cite{GlendinningSidorov2015}, Theorem~2.13) and proved that $b-a<1-2a^*$ is sufficient, with a sharp constant
(their Proposition~3.3 and Corollary~3.9), where $a^*=0.412454\ldots$ is the Thue--Morse constant and
\begin{equation}
1-2a^*=\frac14\prod_{j\ge1}\left(1-2^{-2^j}\right)=0.1750919327\ldots.
\label{eq:astar}
\end{equation}
Since $\dim_H \mathcal K=\htop/\ln2$ for closed invariant sets of the doubling map (\cite{Furstenberg1967}, Proposition~III.1;
\cite{LabarcaMoreira2006}),
the zero-entropy region is
bounded by $\wc(q_c)$, which we obtain by solving $b=\chi(a)$, through lemma~\ref{lem:reduction} when the hole
contains $0$. The resulting curve (figure~\ref{fig:cl}(a)) is piecewise linear with slopes $\pm2$ on the plateaus of
$\chi$ plus a devil's-staircase-like exceptional set. We checked it with two subshifts of finite type (SFTs) that bracket the
entropy: $\mathrm{SFT}^{\rm in}_m$ allows the binary words of length $m$ whose dyadic cylinder lies in the surviving
set, and $\mathrm{SFT}^{\rm out}_m$ those whose cylinder meets it, so that
$\htop(\mathrm{SFT}^{\rm in}_m)\le\htop\le\htop(\mathrm{SFT}^{\rm out}_m)$. At $51$ values of $q_c$ the bracket never
contradicts the curve. Figure~\ref{fig:cl}(a) shows in colour the upper bound, the entropy of
$\mathrm{SFT}^{\rm out}_{18}$, on a grid of windows; it vanishes at every point below the curve, except at three
points very close to it, where it vanishes from $m=20$ on.

For $q_c=1/2$ the boundary is attained at $a=a^*$, where the dimension is zero. This gives
\begin{equation}
\wc=4a^*-1=0.64981613456043\ldots
\label{eq:wc}
\end{equation}
and $\htop=0$ for $w\le\wc$, $\htop>0$ for $w>\wc$. That the entropy vanishes at $\wc$ itself follows from
Theorem~4.6 of \cite{GlendinningSidorov2015}: for the sequence of rationals $(1/2,1/2,\dots)$ their hole is
$(a^*,\chi(a^*))=(a^*,1-a^*)$ (their Proposition~3.3), and $\mathcal K(a^*,1-a^*)$ is uncountable of zero Hausdorff dimension; lemma~\ref{lem:reduction}
transfers this to our window.
An SFT bracket with $m=22$ locates the threshold independently, to six digits.

\subsection{The Thue--Morse cascade}\label{sec:cascade}

For $q_c=1/2$, a periodic point survives iff its whole orbit stays at distance at least $h$ from the integers.
For every period $L\le32$ we computed $h_L=\max_{\rm cycles}\min_t\|2^tx\|$ over the cycles of smallest period $L$
($x=r/(2^L-1)$ with $r$ an integer). We used a depth-first search over binary words that discards a prefix as soon
as one of its suffixes cannot lie in $[h,1-h]$, in exact integer arithmetic; for $L\le24$ it agrees with a plain
enumeration of all cycles.
Writing $t_j$ for the Thue--Morse sequence \cite{AlloucheShallit1999}, the cycles of period $2^\nu$ that attain the largest $h$ are the
Thue--Morse cycles $01,\,0110,\,01101001,\dots$, with
\begin{equation}
h_\nu=\frac14\,\frac{\prod_{j=1}^{\nu-1}\left(1-2^{-2^j}\right)}{1-2^{-2^\nu}},
\quad w_\nu=1-2h_\nu,
\label{eq:hk}
\end{equation}
that is $w_\nu=1/3,\,3/5,\,11/17,\,167/257,\dots$, with Fermat denominators. Equation~\eqref{eq:hk} is the case of the sequence $(1/2,1/2,\dots)$ in Lemma~3.2 of \cite{GlendinningSidorov2015},
and \.Zyczkowski and Bollt obtained the same values for the tent map with a symmetric gap \cite{ZyczkowskiBollt1999}.
They also follow from Proposition~2.16 of \cite{ACS2009}, which gives for every
period $2^k(2l+1)$ the extremal periodic sequence in terms of Thue--Morse blocks. We verified in exact rational
arithmetic that the Thue--Morse cycle of period $2^\nu$ attains equation~\eqref{eq:hk} for $\nu\le6$, and by the search
above that it is the best cycle of its period for $2^\nu\le32$.

Dividing equation~\eqref{eq:hk} by equation~\eqref{eq:astar} gives
\begin{equation}
h_\nu=\frac{1-2a^*}{\left(1-2^{-2^\nu}\right)^2\prod_{j>\nu}\left(1-2^{-2^j}\right)},\qquad
\wc-w_\nu=4(1-2a^*)\,2^{-2^\nu}\left[1+2^{1-2^\nu}+O\!\left(2^{-2^{\nu+1}}\right)\right],
\label{eq:dexp}
\end{equation}
which we checked numerically, including the correction term. The convergence is doubly exponential, in contrast with the geometric
(Feigenbaum) accumulation of cascades in unimodal maps, as \.Zyczkowski and Bollt noted \cite{ZyczkowskiBollt1999},
although the kneading sequence at the Feigenbaum point is itself an automatic sequence closely related to the
Thue--Morse sequence \cite{AlloucheCosnard1983}.
The prefactor and the correction in equation~\eqref{eq:dexp} follow directly from the product formula; we did not
find them stated in \cite{ZyczkowskiBollt1999,GlendinningSidorov2015,ACS2009,Nilsson2010}.

For all other periods, $L=3,5,6,7,\dots$, Theorem~1.3 of \cite{ACS2009} places the thresholds beyond all powers of
two, so $h_L<1-2a^*$; our search confirms it for $L\le32$. These cycles appear only after $\wc$, together with the positive entropy. Therefore,
\emph{below $\wc$ the only cycles in the survivor set are the Thue--Morse cycles}. We confirmed for $L\le32$ that the
number of fixed points of $\sigma^L$ in $\mathcal K$ for $w<\wc$ is at most the number of points on the Thue--Morse
cycles whose period divides $L$, and that above $\wc$ it grows with $L$. Glendinning and Sidorov showed that the doubling map with a hole has routes to chaos
through arbitrary sequences of products of natural numbers, which generalise period doubling
\cite{GlendinningSidorov2015}; the symmetric hole follows the period-doubling route.

\subsection{Entropy above \texorpdfstring{$\wc$}{w\_c}}\label{sec:entropy}

For $h=(1-w)/2$ let $X_h$ be the subshift of binary sequences all of whose shifts lie lexicographically in
$[h,1-h]$; $\htop$ is its entropy. As a set of points, $X_h$ is the set
$F(h)=\{x:\,\|2^tx\|\ge h,\ t\ge0\}$ of dyadically badly approximable numbers studied by Nilsson
\cite{Nilsson2010}, and $\dim_H F(h)=\htop/\ln2$. The dimension is continuous in $h$ \cite{LabarcaMoreira2006,Nilsson2010},
constant on intervals whose union has full Lebesgue measure, and zero exactly for $h\ge1-2a^*$
\cite{Moreira2001,Nilsson2010}. Nilsson also showed that a substitution built on the Thue--Morse words relates the
dimension on successive ranges of $h$ and divides it by a power of two \cite{Nilsson2010}. Applied to the word $001$,
whose interval has $\dim_H F=\log_2\varphi$, this gives two plateaus close to $\wc$ in closed form:
\begin{equation}
\htop=\tfrac12\ln\varphi\quad\text{for}\quad\tfrac{41}{63}\le w\le\tfrac{667}{1008},
\label{eq:plateau}
\end{equation}
and
\begin{equation}
\htop=\tfrac14\ln\varphi\quad\text{for}\quad \tfrac{887}{1365}\le w\le\tfrac{908527}{1397760},
\label{eq:plateau2}
\end{equation}
the second starting $7.2\times10^{-7}$ above $\wc$; in general $\htop=\ln\varphi/2^{j}$ on the $j$-th interval of
the chain. Further up, a wide plateau visible in figure~\ref{fig:cl}(b) is
\begin{equation}
\htop=\ln\theta=0.414\quad\text{for}\quad\tfrac{21}{31}\le w\le\tfrac{173}{248},
\label{eq:plateau3}
\end{equation}
with $\theta=1.5129$ the largest root of $x^4-x^3-x^2+x-1$. Its lower edge, $21/31=1-2\cdot5/31$, is the appearance
threshold of the period-5 cycle $00101$, and the chain of this word gives the plateaus $\tfrac12\ln\theta$ and $\tfrac14\ln\theta$
(appendix~\ref{app:cert}), the second of which lies between $\wc$ and the interval of
equation~\eqref{eq:plateau2}. \.Zyczkowski and Bollt found the plateaus $\ln\theta$
and $\tfrac12\ln\varphi$ numerically \cite{ZyczkowskiBollt1999}, and the edges also follow from the classification of
entropy plateaus for unique expansions in non-integer bases by Alcaraz Barrera, Baker and Kong
\cite{AlcarazBakerKong}. Between $\wc$ and each of these intervals lie infinitely many other plateaus. We certified
the five plateaus in exact integer arithmetic, with a bracket between inner and outer SFTs; the words that define
them, the theorems they rest on and the certificates are given in appendix~\ref{app:cert}.

Along these chains the plateaus approach $\wc$ in a regular way. Let $w_*$ be the lower edge of a plateau. Then $\htop\log_2[1/(w_*-\wc)]$ is close to $2.5$ for the four plateaus of the two chains ($2.1$
for $\ln\theta$): from one step of the chain
to the next the entropy halves and $\log_2[1/(w_*-\wc)]$ roughly doubles. These five plateaus also bound the
entropy between them. A wider window has a larger survivor set, so $\htop$ does not decrease with $w$, and between
two consecutive plateaus it lies between their values. Taking the edges of the five intervals and the gaps between
them, we obtain
\begin{equation}
\frac{1.24}{\log_2[1/(w-\wc)]}\le\htop(w)\le\frac{2.65}{\log_2[1/(w-\wc)]},\qquad
4.45\times10^{-8}\le w-\wc\le0.048,
\label{eq:hstair}
\end{equation}
where the lower end of the range is the lower edge of the plateau $\tfrac14\ln\theta$, the threshold of the cycle
of period $20$ that defines it, and the upper end is the upper edge of the plateau $\ln\theta$. Both constants come
from the wide gap between the plateaus $\tfrac12\ln\varphi$ and $\ln\theta$. A window one lattice site above $\wc$ lies in this range for
$N\le2^{24}$, so its entropy lies between $1.24/\log_2N$ and $2.65/\log_2N$.
Closer to $\wc$ we expect the same law, since every further step of the chain halves the entropy and roughly doubles
$\log_2[1/(w-\wc)]$, so that the entropy of a window one lattice site above $\wc$ vanishes only logarithmically as $N\to\infty$. Of
course, beyond the range of equation~\eqref{eq:hstair} this is an observation; we have no proof of it.

Two classical facts about lattice windows are used in section~\ref{sec:quantum}. A window whose edges lie at
multiples of $1/N$ depends only on blocks of $n$ symbols, so its survivor set is a subshift of finite type of memory
$n$ and the inner SFT of depth $m\ge n$ gives its $\htop$ exactly; we computed it for every lattice window used in
section~\ref{sec:quantum}. The potential $-\tfrac12\ln2$ is constant, so the topological pressure
is $P(1/2)=\htop-\tfrac12\ln2$ and $\ee^{P(1/2)}=\ee^{\htop}/\sqrt2$. The two classical values used in
section~\ref{sec:quantum} follow from the same count: about $\ee^{\htop t}$ periodic orbits of period $t$ survive,
each stretched by $2^t$. Adding their weights $2^{-t}$ gives the classical survival probability $\ee^{-\gamma t}$,
with escape rate $\gamma=\ln2-\htop$; adding their semiclassical amplitudes $2^{-t/2}$ gives $\ee^{P(1/2)t}$. In the
semiclassical theory of chaotic scattering
$P(1/2)$ sets the decay of the longest-lived resonances, which is slower than the classical decay when the repeller is
chaotic and equal to it when it is a single periodic orbit \cite{GaspardRice1989}; for the single-orbit multiplets
of the open baker map the modulus $2^{-1/2}$ saturates this value \cite{Ermann2026PRL}.

\section{Quantum results}\label{sec:quantum}

\subsection{Multiplets below \texorpdfstring{$\wc$}{w\_c}}\label{sec:mult}

As for the single orbits of \cite{Ermann2026PRL}, a cycle $\mathcal O$ of period $T$ inside the window gives a multiplet of $T$ resonances close to
\begin{equation}
\lambda_{\mathcal O,k}=2^{-1/2}\,\ee^{\,\mathrm i(\Phi_{\mathcal O}+2\pi k)/T},\quad k=0,\dots,T-1,
\label{eq:mult}
\end{equation}
where $\Phi_{\mathcal O}=-2\pi N S_{\mathcal O}$ and $S_{\mathcal O}=-\sum_{\epsilon_t=1}q_t$ is the classical action
of the orbit, computed exactly from the symbolic orbit \cite{Ermann2026PRL}; the sum runs over the points $q_t$ of the
orbit whose symbol $\epsilon_t=\lfloor2q_t\rfloor$ is $1$. The multiplet is the open-map counterpart of the resonances generated by a single hyperbolic orbit
\cite{GerardSjostrand1987}, of the periodic-orbit modes of the closed baker map \cite{Ermann2008POM} and of the
resonances built from short periodic orbits \cite{Novaes2009,Pedrosa2012}, and, since here the trapped set contains
only a few cycles, it is the limiting case of the scarring of long-lived resonances by the shortest periodic orbits of the
repeller \cite{ErmannCarloSaraceno2009}. The modulus $2^{-1/2}$ is set by the
stretching, which is the same for all cycles, and the phases by the actions. In the Walsh quantisation the multiplet
is exact, and its phases are fixed by the product of the Hadamard signs along the cycle instead of the action
(section~\ref{sec:model}). We measure the agreement by the largest distance $\mathrm{err}_T$ between a member of
equation~\eqref{eq:mult} and the computed eigenvalue assigned to it, with an optimal one-to-one assignment of all the
predicted members to the computed eigenvalues. For comparison, neighbouring members of a multiplet are
$2^{1/2}\sin(\pi/T)$ apart, $0.54$ for $T=8$ and $0.28$ for $T=16$. The multiplets of all the surviving cycles are
present at once in the spectrum (for the 2- and 4-cycles this was shown in \cite{Ermann2026PRL}): at $w=0.645$ and $n=11,12$ the computed 2- and 4-multiplets lie within $0.015$ of
equation~\eqref{eq:mult}, and at $w=0.6485$ the 8-multiplet lies within $0.045$, although the 8-cycle lies only one
to three sites inside the window. Even and odd $M$ behave alike. As a rule, a cycle is resolved when its extreme
points lie about two or three sites inside the edges of $W$; closer to the edge its multiplet can still be identified,
with errors of a few hundredths. Even in the widest window below $\wc$ the extreme points lie only $N(\wc-w_\nu)/2$
sites inside, so the cycle of period $2^\nu$ needs $N(\wc-w_\nu)/2\gtrsim2.5$, that is $N\gtrsim7\cdot2^{2^\nu}$ by
equation~\eqref{eq:dexp}. Therefore the cascade is cut off quantum mechanically at $\nu_{\max}\simeq\log_2\log_2N$. In terms of time the criterion reads
$2^\nu\lesssim n-2.8$: a cycle is resolved when its period is shorter than $n=\log_2N$, twice the Ehrenfest time $t_E=\frac12\log_2N$ in
which a minimal wave packet is stretched over the torus \cite{BermanZaslavsky1978}, minus about three steps. This is natural on the lattice: a lattice window is a
subshift of finite type of memory $n$ (section~\ref{sec:entropy}), and the threshold of the cycle of period $2^\nu$ lies at a distance of order $2^{-2^\nu}$
from $\wc$, which such a window resolves only if $2^\nu\lesssim n$. The period-locking plateaus of
\cite{Ermann2026PRL} are resolved by a criterion of the same kind, up to periods of about $n-2$.

\begin{figure}
\centering
\includegraphics[width=\textwidth]{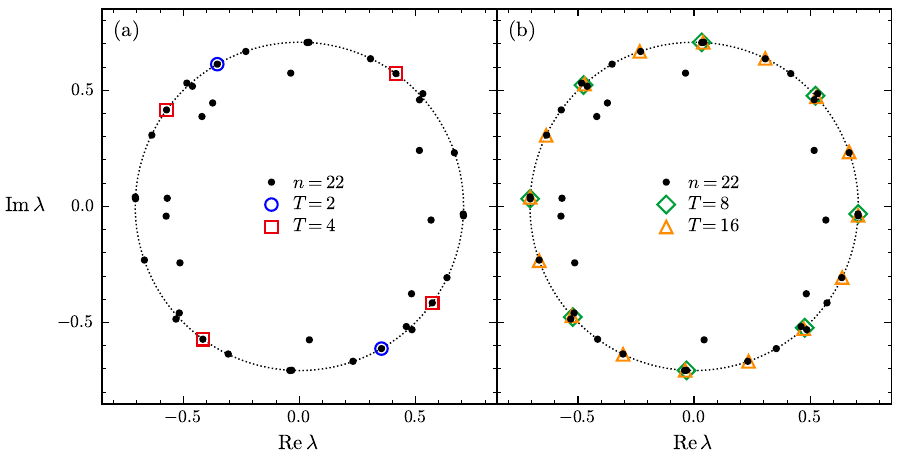}
\caption{Multiplets below the threshold at $q_c=1/2$ (matrix-free Arnoldi). Leading $40$ eigenvalues (dots) for $n=22$
with $w$ midway between $w_4$ and $\wc$, and the predicted multiplets of equation~\eqref{eq:mult} for the cycles of
periods $T=2,4$ (a) and $T=8,16$ (b) (open symbols). The dotted circle is $|\lambda|=2^{-1/2}$. The other ten
eigenvalues are the eigenvalue $\lambda_{\rm D}$ discussed below, at $|\lambda|=0.612$, and the top of the remaining
spectrum, $|\lambda|\le0.58$.}
\label{fig:mult}
\end{figure}

The next step, the 16-cycle, appears at $w_4=167/257$, only $1.07\times10^{-5}$ below $\wc$, and by this estimate
needs $N\gtrsim5\times10^5$. With the Arnoldi method we computed the $36$--$40$ leading eigenvalues in the window
halfway between $w_4$ and $\wc$ for $N=2^{18}$ to $2^{25}$, and in further windows with $M$ in $[Nw_4,N\wc]$ for
$N=2^{20}$ to $2^{23}$. All $2+4+8+16=30$ predicted resonances are found (figure~\ref{fig:mult}). The 16-multiplet is
already identified at $N=2^{18}$, where the 16-cycle lies less than one site inside the window
($\mathrm{err}_{16}=0.037$), and its error decreases as the window edges move away from the threshold: it is below
$0.004$ at $N=2^{23}$ and $0.0005$ at $N=2^{25}$, where the window ends $90$ sites per side below $N\wc$. This is the fourth step of the cascade resolved in the quantum spectrum.

\begin{figure}
\centering
\includegraphics[width=0.6\textwidth]{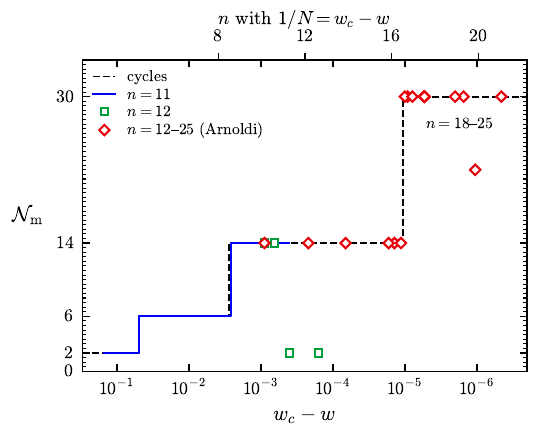}
\caption{The cascade in the quantum spectrum at $q_c=1/2$. Number $\mathcal N_{\rm m}$ of resolved multiplet resonances,
the sum of the periods of the Thue--Morse cycles inside the window whose multiplet agrees with
equation~\eqref{eq:mult} within $\mathrm{err}_T\le0.05$, versus the distance $\wc-w$ to the entropy threshold, for $n=11$ (line) and $n=12$ (squares),
dense diagonalisation, and $n=12$--$25$ (diamonds, Arnoldi). The windows halfway between $w_4$ and $\wc$ at $n=18$--$25$ have the
same $w$ ($M/N=0.6498108$) and fall on one point (labelled); at all eight sizes the $30$ resonances are resolved.
Dashed line: the number of points on the surviving cycles, $2$, $6$, $14$ and $30$. Top axis: the value of $n$ for
which one lattice site equals $\wc-w$; the axis ends just beyond the window closest to $\wc$ with $40$ eigenvalues. The three points below
the dashed line are windows within five sites of $\wc$, where a multiplet exceeds the tolerance (see text).}
\label{fig:route}
\end{figure}

Figure~\ref{fig:route} summarises the route as the quantum spectrum sees it. The number of resolved multiplet
resonances, counted over the multiplets with $\mathrm{err}_T\le0.05$, steps from $2$ to $6$, $14$ and $30$ at the
thresholds of the cycles, as does the number of points on the surviving cycles, and each step appears only at the
sizes that resolve the cycle. The exceptions are three windows within five sites of $\wc$, two at $n=12$ and one at
$n=22$, where a multiplet is perturbed beyond the tolerance (section~\ref{sec:hybrid}).

Besides the multiplets the spectrum contains one eigenvalue, $\lambda_{\rm D}$, that changes little with $N$: the
eigenvalue at $|\lambda|=0.612$ in figure~\ref{fig:mult}. It lies close to the leading eigenvalue of an
$N$-independent operator $\mathcal D$ that describes the map near its discontinuity $p=0$, in lattice units of
momentum (appendix~\ref{app:D}): in the Arnoldi windows at $q_c=1/2$ that lie more than three sites per side below
the threshold ($n=19$--$25$) the block has an eigenvalue within $6\times10^{-3}$ of it. It has no counterpart in the Walsh quantisation, where the nonzero
spectrum of a window that traps a single cycle is the multiplet alone (section~\ref{sec:model}).

\subsection{Hybridisation near \texorpdfstring{$\wc$}{w\_c}: a quantum precursor of the transition}\label{sec:hybrid}

\begin{figure}
\centering
\includegraphics[width=\textwidth]{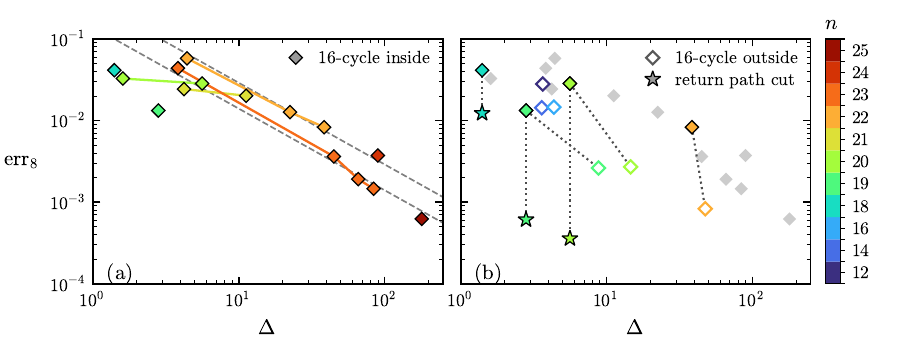}
\caption{Quantum precursor of the transition. Maximum error $\mathrm{err}_8$ of the 8-multiplet,
equation~\eqref{eq:mult} (section~\ref{sec:mult}), versus the distance $\Delta=N\wc-M$ of the window to the entropy
threshold, in lattice sites, at $q_c=1/2$ (Arnoldi, $n=12$--$25$). The colour gives $n$ (colour bar, one step per size
computed). (a) Windows with the 16-cycle inside; thin lines join the windows of the same $n$. Dashed
lines: $0.29/\Delta$ and $0.14/\Delta$, the means of $\mathrm{err}_8\,\Delta$ at $n=22$ and $23$. (b) Controls. Open
diamonds: windows without the 16-cycle ($M<Nw_4$); at $n=19$, $20$ and $22$ a dotted segment joins each of them to the
last window that contains the 16-cycle (filled diamond). Stars ($n=18$, $19$, $20$): the window of the filled diamond
above it, joined to it by a vertical dotted segment, with the 16-cycle still inside but with the return path from the
16- to the 8-cycle cut. On the lattice graph this path opens only when the window is enlarged by
$\lceil\Delta/2\rceil$ sites per side, that is, at $\wc$. We take the shortest such paths in the enlarged window and
remove from the window those of their sites that lie at least $20$ sites away from every cycle point. These sites lie
thousands of sites from the window edges, and none of them is on a shortest forward path $8\to16$, so $\Delta$, the
cycles and the forward connection do not change (section~\ref{sec:hybrid}). Grey: the other windows of (a).}
\label{fig:prec}
\end{figure}

Close to the threshold the multiplets are no longer independent. Within a few tens of sites of $\wc$, with the
16-cycle inside the window, the 8-multiplet is perturbed by up to $0.06$, although the 8-cycle lies more than $350$
sites inside the window. The perturbation acts on one of the two parity sectors. For even $M$ the reflection
$j\to N-1-j$ commutes with the propagator and acts on the 8-cycle as a shift by half a period, so each sector contains
alternate members of the multiplet, and since only one sector is perturbed the moduli alternate with $k$. In figure~\ref{fig:prec} we show the error of the 8-multiplet against the distance
$\Delta=N\wc-M$ of the window to the threshold, in lattice sites. Close to the threshold, $\Delta\lesssim5$, it is
$0.01$--$0.06$ with or without the 16-cycle (the windows without it, at $n=12,14,16$, have the 16-cycle only about
two sites outside the window edge). Further away the two cases separate: without the 16-cycle the error drops to a
few $10^{-3}$, while with it the error decreases slowly, roughly as $1/\Delta$, and reaches $6\times10^{-4}$ only at
$\Delta\approx180$ ($n=25$). The cleanest comparison is at fixed $n$: shrinking the window until the 16-cycle has
just left it lowers the error by a factor $5$--$10$ at $n=19$, $20$ and $22$, while $\Delta$ grows
only by a factor $1.2$--$3$, which in a $1/\Delta$ law would lower the error only by that factor
[figure~\ref{fig:prec}(b)]. The perturbation therefore needs the 16-cycle inside the window, or within a few sites of
its edge, and the edge close to the threshold.

The lattice graph, the eigenvectors and the phases show where the perturbation comes from. On the lattice graph of
the window (section~\ref{sec:model}) the 8-cycle is joined to the 16-cycle by a one-way chain of $n-8$ sites, and the
4-cycle is joined to both; no chain goes back. In the Walsh quantisation the propagator is then block triangular, and
its nonzero spectrum is exactly the union of the multiplets. In the Fourier quantisation we computed right
eigenvectors in nine windows at $n=18$--$22$, all with even $M$, and we measure the weight of an eigenvector on a
cycle by $\sum|\langle q,p|\psi^R\rangle|^2/\|\psi^R\|^2$ over symmetric coherent states $|q,p\rangle$ at the points
of the cycle. The right eigenvectors of the perturbed sector of the 8-multiplet carry less weight on the 8-cycle than
those of the other sector, by a factor $2$--$10$, and, in six of the seven windows in which a single sector is
perturbed, relatively more on the 16-cycle. These eigenvectors are hybridised with the 16-cycle along the connection
$8\to16$. The right eigenvectors of the 16-multiplet carry almost no weight on the 8-cycle (below $1.4\times10^{-3}$
for $n\ge19$), so the coupling has the direction of the classical connection.

The phases select the sector that hybridises. Members of different parity do not mix, and within one parity sector
the members of a $T$-multiplet are $4\pi/T$ apart in phase, so the detunings $\delta_\pm$ of the two sectors of the
8-multiplet from the nearest member of the 16-multiplet of the same parity satisfy $\delta_++\delta_-=\pi/8$. In the
six windows at $n=18,\,19,\,21$ and $22$ the perturbed sector is the closer one ($\delta\le0.031\pi$); at $n=20$ the
two detunings are nearly equal and both sectors are perturbed close to the threshold. Each member of the closer sector
has one partner in the 16-multiplet at phase distance $\delta$; the four partners are $\pi/2$ apart and form the
perturbed class of $k$ mod $4$, which therefore depends on $N$ through the action phases.

A near-resonance moves the eigenvalues only if amplitude can return from the 16-cycle to the 8-cycle: with a
coupling in one direction only, the propagator is block triangular, as in the Walsh quantisation, and the multiplets
do not move. On the lattice
graph this return path is closed, but it opens as soon as the window is enlarged by $\lceil\Delta/2\rceil$ sites on
each side, that is, exactly at the threshold $N\wc$, in all nine windows: the two-way connection between the 8- and
16-cycles appears together with the positive entropy. The Fourier propagator has tails that decay as $1/(\pi d)$
with the distance $d$ in sites, so a return path blocked by about $\Delta/2$ sites is crossed with an amplitude of
order $1/\Delta$. The measured shifts decrease in nearly the same way within each size: the error of the 8-multiplet
decreases as $\Delta^{-0.9}$ at $n=22$ and as $\Delta^{-1.1}$ at $n=23$, close to the
dashed lines of figure~\ref{fig:prec}(a), which are proportional to $1/\Delta$. The prefactor changes from one size
to the next, as the detunings do. We have not derived this law. For two levels at detuning $\delta$, with a forward
coupling $a$ that does not depend on $\Delta$ and a reverse coupling $b$ of order $1/\Delta$, the eigenvalues move by $(\delta^2/4+ab)^{1/2}-\delta/2$
\cite{ErmannCarlo2026}, which is $ab/\delta\propto1/\Delta$ only for shifts small compared with $\delta$ and
$(ab)^{1/2}\propto\Delta^{-1/2}$ otherwise. At $n=22$ the three windows have the same detuning and the shifts exceed
it, so there the two-level model gives $\Delta^{-1/2}$ and not the measured law.

We tested this reading directly, by cutting the return path while keeping the 16-cycle inside the window. In the
window enlarged by $\lceil\Delta/2\rceil$ sites per side we took all the shortest paths from the 16- to the 8-cycle on
the lattice graph, of length $n-8$ like the forward chain. Their sites inside the window, other than the cycle sites,
lie thousands of sites away from the window edges, and none of them lies on a shortest path of the forward connection
$8\to16$. We removed from the window either the far part of these paths, the sites at least $20$ sites away from
every cycle point, or, as a control, holes next to the 16-cycle that are not on the paths (the mirror images, with
respect to the cycle points, of the path sites within $20$ sites of them). At $n=19$ and $20$ ($\Delta=2.8$ and
$5.6$), removing the far part lowers $\mathrm{err}_8$ from $0.013$ and $0.029$ to $0.0006$ and $0.0004$, below the
level of the windows without the 16-cycle [figure~\ref{fig:prec}(b)], and leaves the 16-multiplet as it was. The control
holes perturb the 16-multiplet but leave $\mathrm{err}_8$ at $0.021$--$0.022$, and the cut sites displaced along the
window do not lower it either. The perturbation therefore travels along the return path; it is not a detuning of the
near-resonance by the window. At $n=18$ and $\Delta=1.4$ the test is less clean ($\mathrm{err}_8$ goes from $0.042$
to $0.012$ with the cut and to $0.026$ with the control): at such distances the windows have errors of
$0.01$--$0.06$ even without the 16-cycle.

The 4-cycle gives a further check. Its return path opens only at $w=13/20$, $1.8\times10^{-4}$ above $\wc$. For $n\ge18$
this is $24$ sites or more per side beyond the threshold, and the 4-multiplet is not perturbed (errors below
$0.004$), although its near-resonant eigenvectors also extend to the 16-cycle. At smaller sizes the path opens only a few
sites beyond $\wc$, and the 4-multiplet is perturbed near $\wc$, with errors of $0.02$--$0.07$ at $n=11$--$16$
(figure~\ref{fig:route}).

Since the return path from the 16- to the 8-cycle is closed in the classical window, the amplitude crosses the missing sites through the tails of the Fourier
propagator, that is, by diffraction. The perturbation of the 8-multiplet is thus a quantum precursor of the
transition: through diffraction, the spectrum is sensitive to a classical connection that exists only above $\wc$.
Of course, at a fixed $w<\wc$ the distance $\Delta=N(\wc-w)$ grows with $N$, so the precursor is a finite-$N$ effect
that disappears as $N\to\infty$.

In the open three-disk system the interaction of several periodic orbits is known to shape the resonance chains
\cite{Weich2014}. The coupling of periodic orbits linked in one direction, its size and its relation to the window
edges and to the discontinuity of the map at $q=1/2$ are studied for repellers made of a few orbits in
\cite{ErmannCarlo2026}.

\subsection{The jump of \texorpdfstring{$|\lambda_0|$ at $\wc$}{|lambda0| at w\_c}}\label{sec:jump}

\begin{figure}
\centering
\includegraphics[width=\textwidth]{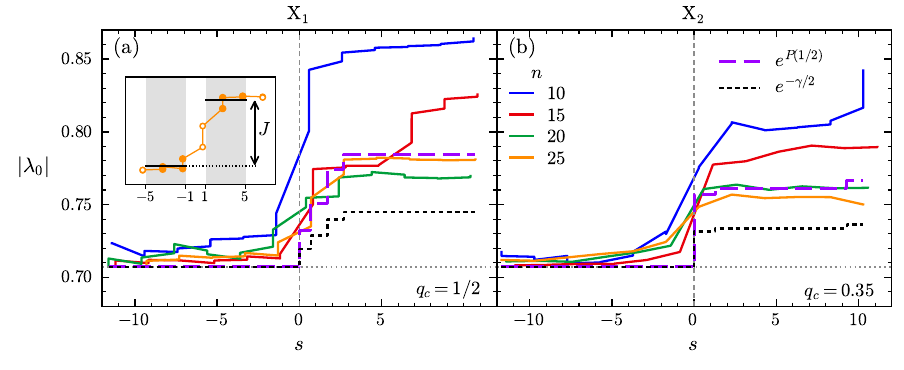}
\caption{Leading resonance near the threshold (matrix-free Arnoldi). (a) Modulus of the leading resonance
$|\lambda_0|$ versus the distance to the threshold in lattice sites, $s=M-N\wc(q_{\rm eff})$ (section~\ref{sec:jump}),
for $q_c=1/2$ and $n=10,15,20,25$. The two window sizes with the same $s$
(section~\ref{sec:jump}) are joined by a vertical segment. Long-dashed purple line: pressure value $\ee^{P(1/2)}=\ee^{\htop}/\sqrt2$; short-dashed black line: $\ee^{-\gamma/2}$ with
$\gamma=\ln2-\htop$ the classical escape rate, both drawn as steps through the exact $\htop$ of each lattice window
at $n=25$ (the step from $\htop=0$ at $s=0$; for $s<0$, where $\htop=0$, both equal $2^{-1/2}$) (section~\ref{sec:entropy}). Dotted line: $2^{-1/2}$; vertical dashed line: $s=0$, the
crossing X$_1$ of figure~\ref{fig:cl}(a). Inset: definition of the jump $J$ at $n=25$. Circles are the windows, two for each value of $s$; $J$ is the
difference between the mean of $|\lambda_0|$ over the four windows with $1\le s\le5$ and over the four with
$-5\le s\le-1$ (filled circles in the shaded bands, means as horizontal segments; table~\ref{tab:jump}). (b) The same for $q_c=0.35$ (crossing X$_2$), where the
vertical segments are shorter than $2\times10^{-3}$ for $n\ge11$, since X$_2$ lies on a flank of the boundary.}
\label{fig:q}
\end{figure}

Below $\wc$ the leading resonance is close to $2^{-1/2}$. A few sites below the threshold it exceeds $2^{-1/2}$ by up to
a few hundredths, and at a fixed distance in sites this excess does not decrease with $N$ up to $N=2^{25}$
(section~\ref{sec:pressure}). Further below, the excess decreases with $N$ (at $q_c=1/2$ the mean $|\lambda_0|$ of the
windows more than $12$ sites below $N\wc$ is $0.718$ at $n=10$ and at most $0.709$ for $n\ge17$).
In figure~\ref{fig:q}(a),(b) we show $|\lambda_0|$ across the threshold. Above $\wc$ the leading
resonance jumps. To follow the jump to large $N$ we computed the six leading eigenvalues with
the Arnoldi method for $N=2^{10}$ to $2^{25}$, with more than twenty window sizes $M$ per $N$ near $\wc$, for
$q_c=1/2,\,0.35,\,0.30$, which cross the threshold at the points X$_1$, X$_2$, X$_3$ of figure~\ref{fig:cl}(a). The
centre of the window is $q_{\rm eff}=j_c/N$ for even $M$ and $j_c/N+1/(2N)$ for odd $M$, and the coordinate
$s=M-N\wc(q_{\rm eff})$ is the distance to the threshold in lattice sites (for even $M$, $s=-\Delta$ of
section~\ref{sec:hybrid}). At $q_c=1/2$ and $0.35$ the boundary has slope $\pm2$ at the
crossing, so each added site, which moves the centre by $1/(2N)$, moves $N\wc(q_{\rm eff})$ by one site. Every second
added site moves it in the same direction as $M$ and leaves $s$ unchanged, so $s$ takes the values $s_0+2m$, each
realised by two consecutive sizes that differ by one site at one edge: an odd $M$ and $M+1$ at $q_c=1/2$ (site added on
the left), an even $M$ and $M+1$ at $q_c=0.35$ (site added on the right). At $q_c=1/2$ both edges are critical, and
the $|\lambda_0|$ of the two sizes differ, by $2\times10^{-3}$ on median and by up to $0.04$ just above the threshold
(vertical segments in figure~\ref{fig:q}(a)). At $q_c=0.35$ the right edge is not critical, and the two values agree
within $2\times10^{-3}$ for $n\ge11$ (figure~\ref{fig:q}(b)). We define the jump $J$ as the
mean of $|\lambda_0|$ over $1\le s\le5$ minus the mean over $-5\le s\le-1$, four windows on each side, and give as its
error the standard error of the two means. The windows are deterministic, so this error measures the spread between
windows (table~\ref{tab:jump} for $q_c=1/2$).

\begin{table}
\caption{Jump of the leading resonance across $\wc$ at $q_c=1/2$ (Arnoldi). Columns: mean $|\lambda_0|$ below
($-5\le s\le-1$) and above ($1\le s\le5$), their difference $J$ (in parentheses, the standard error from the spread of the four
windows on each side), the classical value $J_{\rm cl}$ of
equation~\eqref{eq:jcl} and its pressure analogue $J_P$, equation~\eqref{eq:jp}, both computed for the same lattice
windows, and the midpoint of the largest step between consecutive $M$.}
\label{tab:jump}
\centering
\begin{tabular}{ccccccc}
\toprule
$n$ & below & above & $J$ & $J_{\rm cl}$ & $J_P$ & largest step at $s=$ \\
\midrule
10 & 0.732 & 0.854 & 0.122(5) & 0.089 & 0.189 & $-0.41$ \\
11 & 0.719 & 0.822 & 0.103(8) & 0.078 & 0.164 & $+0.18$ \\
12 & 0.732 & 0.834 & 0.103(14) & 0.080 & 0.168 & $+1.35$ \\
13 & 0.720 & 0.834 & 0.114(7) & 0.075 & 0.157 & $+0.71$ \\
14 & 0.720 & 0.798 & 0.078(17) & 0.050 & 0.105 & $+1.41$ \\
15 & 0.714 & 0.777 & 0.063(1) & 0.044 & 0.090 & $-0.18$ \\
16 & 0.721 & 0.769 & 0.048(4) & 0.042 & 0.086 & $-0.35$ \\
17 & 0.711 & 0.791 & 0.080(5) & 0.042 & 0.086 & $+0.30$ \\
18 & 0.718 & 0.783 & 0.065(4) & 0.044 & 0.090 & $-0.40$ \\
19 & 0.711 & 0.768 & 0.057(2) & 0.042 & 0.086 & $+0.20$ \\
20 & 0.721 & 0.767 & 0.046(6) & 0.043 & 0.089 & $-0.60$ \\
21 & 0.727 & 0.792 & 0.065(5) & 0.043 & 0.089 & $-0.21$ \\
22 & 0.732 & 0.774 & 0.042(6) & 0.038 & 0.078 & $+0.59$ \\
23 & 0.721 & 0.758 & 0.037(3) & 0.035 & 0.073 & $+0.18$ \\
24 & 0.728 & 0.769 & 0.041(4) & 0.036 & 0.075 & $+0.35$ \\
25 & 0.718 & 0.779 & 0.061(4) & 0.036 & 0.075 & $+0.70$ \\
\bottomrule
\end{tabular}
\end{table}

The largest step lies within two sites of $\wc$ for every $n$ and the three window centres. The jump thus sets in
within one or two lattice sites, $\Delta w\sim1/N$, as expected: a window one site above $\wc$ already has an entropy
of order $1/\log_2N$ (section~\ref{sec:entropy}), and the window edge is not sharp on the lattice scale. Non-dyadic sizes
(twice a prime, up to $N=262\,202$) give jumps within $0.005$ of those of the nearest powers of two.

\begin{figure}
\centering
\includegraphics[width=\textwidth]{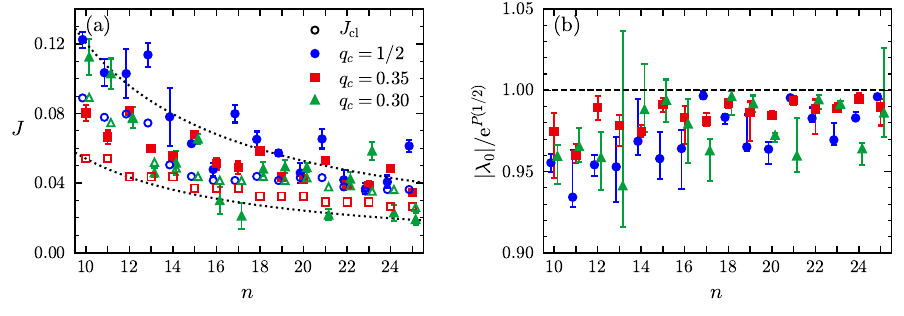}
\caption{Jump and pressure at large $N$ (Arnoldi, $N=2^{10}$--$2^{25}$). (a) Jump $J$ of $|\lambda_0|$ at $\wc$
(section~\ref{sec:jump}) versus $n=\log_2N$ for the three window centres of X$_1$--X$_3$ in figure~\ref{fig:cl}(a)
(filled symbols, with the standard error), and the classical value $J_{\rm cl}$ of equation~\eqref{eq:jcl} for the
same windows (open symbols, same colours). Dotted lines: $(\ee^{\htop/2}-1)/\sqrt2$ with $\htop=1.24/(n-\log_23)$
and $2.65/(n-\log_23)$, the bounds of equation~\eqref{eq:hstair} for $q_c=1/2$ at $s=3$. (b) Ratio $|\lambda_0|/\ee^{P(1/2)}$ above the threshold ($1\le s\le12$):
mean over the windows (symbols) and range (bars: smallest and largest value over the windows, not errors).}
\label{fig:xl}
\end{figure}

The height of the jump decreases with $N$, slowly and not monotonically (figure~\ref{fig:xl}(a)): at $q_c=1/2$ it
goes from $0.12$ at $n=10$ to $0.04$--$0.08$ for $n\ge16$, and for $n\ge15$ it is constant within its scatter,
$0.055\pm0.013$. Most changes from one $n$ to the next exceed the standard error of $J$, and for
$n\ge15$ they are not classical: at $n=15$--$21$ the windows just above the threshold lie mostly on one plateau of
the staircase, where $J_{\rm cl}$ is nearly constant (see below), while $J$ goes from $0.046$ to $0.080$. Both
means contribute to these changes (table~\ref{tab:jump}). Over
$n=10$--$25$ a law $J\propto1/n$ and a power law $J\propto N^{-0.1}$ fit equally well, so the data
do not single out a law.

Over $n=10$--$25$ the decrease has the size of the classical one. With the escape rate $\gamma=\ln2-\htop$ of section~\ref{sec:entropy}, $\ee^{-\gamma/2}$ is the modulus
of an eigenvalue whose state loses its norm at the classical rate, $|\lambda|^{2t}=\ee^{-\gamma t}$; for $144$ windows with $q_c=0.30$--$0.40$, $w>1/2$ and $N\approx2000$,
$\ln|\lambda_0|$ follows $\ln\ee^{-\gamma/2}$ with an rms deviation of $0.034$, mostly a mean offset of $+0.030$. We therefore
compare $J$ with the same average of the classical value,
\begin{equation}
J_{\rm cl}=\langle\ee^{-\gamma/2}\rangle_{1\le s\le5}-\langle\ee^{-\gamma/2}\rangle_{-5\le s\le-1},
\label{eq:jcl}
\end{equation}
where $\htop$ is the exact entropy of each lattice window of the scan (section~\ref{sec:entropy}). Below threshold
$\htop=0$, so the second mean is $2^{-1/2}$. $J_{\rm cl}$ decreases with $N$, but in steps, because $\htop(w)$ is a
staircase, and on average again as $1/n$. This is the law of
equation~\eqref{eq:hstair}: at $q_c=1/2$ and $n\le24$, a window $s\ge1$ sites above the threshold has
$1.24/(n-\log_2s)\le\htop\le2.65/(n-\log_2s)$, hence $J_{\rm cl}=(\ee^{\htop/2}-1)/\sqrt2$ lies between about
$0.44/(n-\log_2s)$ and $0.94/(n-\log_2s)$; the dotted lines of figure~\ref{fig:xl}(a) are these bounds at $s=3$. At
$q_c=1/2$ and $n=15$--$21$ the windows with $1\le s\le5$ lie mostly on the plateau $\tfrac14\ln\varphi$ of
equation~\eqref{eq:plateau2}, where $J_{\rm cl}$ stays at $0.042$--$0.044$. From $n=22$ on they reach the plateau
$\tfrac14\ln\theta$ and the steps closer to $\wc$, and $J_{\rm cl}$ drops to $0.035$--$0.038$; $J$
follows at $n=22$--$24$ but not at $n=25$.

At $q_c=1/2$ and $0.35$ the quantum jump exceeds the classical one at every $n$ (at $q_c=0.30$ the difference changes
sign), and, for $s\ge1$, $|\lambda_0|$ itself lies between $\ee^{-\gamma/2}$ and the pressure value $\ee^{P(1/2)}=\ee^{\htop}/\sqrt2$
(figure~\ref{fig:q}(a); section~\ref{sec:entropy}; for $0<s<1$ see section~\ref{sec:pressure}). The jump defined with the pressure,
\begin{equation}
J_P=\langle\ee^{P(1/2)}\rangle_{1\le s\le5}-2^{-1/2},
\label{eq:jp}
\end{equation}
is about twice $J_{\rm cl}$, because $P(1/2)=\htop-\tfrac12\ln2$ contains $\htop$ with twice the weight it has in
$-\gamma/2$ (table~\ref{tab:jump}). At $q_c=1/2$ and $0.35$ the
quantum jump lies between $J_{\rm cl}$ and $J_P$ for every $n$. It falls short of $J_P$ for two reasons: above the
threshold $|\lambda_0|$ stays slightly below $\ee^{P(1/2)}$, and below it $|\lambda_0|$ exceeds $2^{-1/2}$. At $q_c=1/2$ and $n=17$, $21$ and $25$ the mean ratio $|\lambda_0|/\ee^{P(1/2)}$ above the threshold is
$0.995$--$0.997$ (figure~\ref{fig:xl}(b)) and $J$ is $0.061$--$0.080$. At $n=16$, $20$ and $22$--$24$ the ratio is
$0.96$--$0.98$, the excess below is $0.014$--$0.025$, and $J$ exceeds $J_{\rm cl}$ by only $0.002$--$0.006$, less
than two standard errors.

Since $\htop$ is continuous at $\wc$, the classical part of the jump vanishes as $N\to\infty$, and
equation~\eqref{eq:hstair} suggests that it does so only as $1/\log_2N$. The measured jump also contains the excess
of $|\lambda_0|$ below the threshold, which at $n=15$--$25$ stays at $0.013$ on average while $J_{\rm cl}$ falls to
$0.035$. This excess need not decrease with $N$. The windows of the mean below lie $1$ to $5$ sites below the
threshold at every $N$, and at such distances the multiplets deviate from equation~\eqref{eq:mult} by
$0.01$--$0.06$, of the same order as the excess (section~\ref{sec:hybrid}). At fixed $w<\wc$ the distance in sites
grows as $N(\wc-w)$ and this diffractive perturbation disappears, but at fixed $s$ it need not. The jump at fixed $s$
therefore vanishes if $|\lambda_0|$ keeps following the entropy above the threshold and the excess below it vanishes;
the data up to $N=2^{25}$ support the first condition and do not decide the second.

\subsection{Comparison with the pressure}\label{sec:pressure}

Above threshold ($1\le s\le12$) the ratio $|\lambda_0|/\ee^{P(1/2)}$ lies between $0.93$ and $1$ at $q_c=1/2$ and $0.35$
(figure~\ref{fig:xl}(b)), so $|\lambda_0|$ stays below the pressure value in all these windows. At $q_c=0.30$
the ratio exceeds $1$ in nine windows and reaches $1.036$; these windows lie within seven sites of the threshold,
where $\htop\le0.12$, and $q_c=0.30$ is also the centre with the largest excess of $|\lambda_0|$ below the threshold
(see below). Within one site of the threshold,
$0<s<1$, $|\lambda_0|$ exceeds $\ee^{P(1/2)}$ by up to $0.03$ at $q_c=0.30$ and by up to $0.011$ at
$q_c=1/2$, as expected from a window edge that is not sharp on the lattice scale. The ratio
grows slowly with $N$: at $q_c=1/2$ its mean is $0.95$ for $n=10$--$13$ and $0.98$ for $n=17$--$25$. Of course, a few
sites above the threshold the entropy is small, $\htop\approx0.1$ for $n\ge15$, and there the classical value
$\ee^{-\gamma/2}$ is itself about $0.95$ times the pressure value, so the ratio alone does not separate the two. To
separate them we write $|\lambda_0|=2^{-1/2}\ee^{\beta\htop}$, where $\beta=1/2$ gives the classical decay and
$\beta=1$ the pressure value. Over the windows with $1\le s\le12$ the mean of $\beta$ is $0.81$--$0.86$ at the three
centres. This $\beta$ is measured
from $2^{-1/2}$, which $|\lambda_0|$ does not reach below the threshold. Measured from the jump it is smaller: at
$q_c=1/2$ the means over $n=15$--$25$ are $J=0.055$, $J_{\rm cl}=0.040$ and $J_P=0.083$ (table~\ref{tab:jump}),
which corresponds to $\beta\approx2/3$. The difference between the two values is the excess of $|\lambda_0|$ below
the threshold, discussed next. In both cases the leading resonance grows with the entropy faster than the classical
decay and more slowly than the pressure value.

Below threshold $\htop=0$ and $\ee^{P(1/2)}=2^{-1/2}$, while the mean $|\lambda_0|$ over $-5\le s\le-1$ exceeds
$2^{-1/2}$ by $0.003$--$0.025$ at $q_c=1/2$ and $0.35$ and by up to $0.053$ at $q_c=0.30$
(table~\ref{tab:jump} for $q_c=1/2$). The classical entropy does not account for this excess, which does not decrease
with $N$ up to $N=2^{25}$, and neither does the $N$-independent eigenvalue $\lambda_{\rm D}$ of
section~\ref{sec:mult}, whose modulus near X$_1$, X$_2$ and X$_3$ is $0.612$, $0.610$ and $0.662$, below
$2^{-1/2}$. The excess
is at least in part the precursor of section~\ref{sec:hybrid}: in the seven Arnoldi windows of that section that lie
within six sites below the threshold at $n=18$--$23$, the leading eigenvalue is a member of the perturbed 8- or
16-multiplet.

The Nonnenmacher--Zworski bound \cite{NonnenmacherZworski2009} is an asymptotic statement for
$N\to\infty$, and, to our knowledge, it is not proven for this discontinuous hard-window model. For an open baker
map whose trapped set touches the discontinuities of the map, Nonnenmacher found the spectral radius of the
odd-parity sector at or slightly above the bound \cite{Nonnenmacher2011}. Our data are
consistent with it above threshold, but they neither confirm nor exclude that the excess below threshold vanishes as
$N\to\infty$. For open baker maps of base $b\ge3$ with a smooth cutoff, $N=b^k$ and a trapped set that is an
arithmetic Cantor set of dimension $d$, where the pressure value is $b^{d-1/2}$, Dyatlov and Jin proved that the
spectral radius stays below $b^{-g}$ as $N\to\infty$, with $g>\max(0,\tfrac12-d)$, that is, strictly below the
pressure value \cite{DyatlovJin2017}. Our map is outside this class (base $2$, sharp window, $\mathcal K$ not an
arithmetic Cantor set), and ratios of $0.93$--$1.00$ up to $N=2^{25}$ do not decide whether such an improvement holds
here.

\setcounter{totalnumber}{2}
\subsection{The resonance count above \texorpdfstring{$\wc$}{w\_c}: fractal Weyl law}\label{sec:weyl}

Below $\wc$ the resonances close to modulus $2^{-1/2}$ are counted by the points on the surviving cycles, $2$, $6$,
$14$ and $30$ (figure~\ref{fig:route}), a number that does not grow with $N$. Above $\wc$ the survivor set has
positive entropy, and the last question along the route is whether the number of long-lived resonances then grows as
the fractal Weyl law predicts. The trapped set in the $(q,p)$ plane is
locally the product of two Cantor sets of dimension $\htop/\ln2$ each. Since the stretching is constant, the
Hausdorff, box and information dimensions of each factor coincide, and the law reads
\cite{Sjostrand1990,LuSridharZworski2003,NSZ2014}
\begin{equation}
\mathcal N(c)=\#\{\lambda:\ |\lambda|>c\}\propto N^{d_{\rm cl}},\qquad d_{\rm cl}=\frac{\htop}{\ln2},
\label{eq:weyl}
\end{equation}
for fixed $0<c<|\lambda_0|$. We call $d_{\rm fit}$ the slope of a least-squares fit of $\ln\mathcal N(c)$ versus
$\ln N$, and we use $c=0.2$ unless stated otherwise. Smaller cutoffs are not reliable: the block is strongly
non-normal, and its nearly nilpotent parts produce spurious eigenvalues in double precision, of modulus about
$(10^{-16})^{1/m}$ for a chain of length $m$, which reach $|\lambda|\simeq0.15$ at our largest sizes.

On the route itself the count can be made only at some distance from $\wc$, since close to it the entropy is too
small. At $q_c=1/2$ and $w=0.655$, on the plateau $\htop=\tfrac12\ln\varphi$ of equation~\eqref{eq:plateau}, the law
predicts $d_{\rm cl}=0.347$, and dense diagonalisation at $N=2^{11}$--$2^{14}$ and at sizes twice a prime gives
$d_{\rm fit}=0.33\pm0.02$. This agrees with the law, but one window followed over less than a decade of $N$ is not a
test. We therefore add windows with a wider range of entropies, most of them away from $q_c=1/2$: six one-strip
windows A--F (squares in figure~\ref{fig:cl}(a)) with $\htop$ from $0.20$ to $0.66$, certified at $m=22$ (table~\ref{tab:weyl}), followed up to $N=2^{17}$
(A, B, C) or $2^{16}$ (D, E, F) and at sizes twice a prime from $N=2062$ to $16\,418$.

\begin{table}
\caption{Fractal Weyl exponents for the six windows A--F (squares in figure~\ref{fig:cl}(a)). $d_{\rm cl}=\htop/\ln2$; DFT: least-squares fit of
$\ln\mathcal N(0.2)$ versus $\ln N$ over all dyadic and $2p$ sizes (jackknife error); $V(n)$: exponent of the
number of vertices in non-trivial strongly connected components of the lattice graph, $n=13$--$22$; Walsh: exponent
of the number of Walsh eigenvalues with $|\lambda|>0.2$ (dense diagonalisation at $n=8$--$16$, and, for A--C, of the restriction to the
non-trivial strongly connected components at $n=21,22$; fit error).}
\label{tab:weyl}
\centering
\begin{tabular}{ccccccc}
\toprule
  & $(q_c,w)$ & $\htop$ & $d_{\rm cl}$ & DFT & $V(n)$ & Walsh \\
\midrule
A & $(0.32,0.53)$ & 0.1995 & 0.288 & 0.234(4) & 0.288 & 0.283(12) \\
B & $(0.26,0.535)$ & 0.2812 & 0.406 & 0.382(14) & 0.403 & 0.404(6) \\
C & $(0.45,0.72)$ & 0.4095 & 0.591 & 0.528(3) & 0.591 & 0.580(6) \\
D & $(0.48,0.78)$ & 0.5106 & 0.737 & 0.690(5) & 0.737 & 0.731(3) \\
E & $(0.43,0.87)$ & 0.6094 & 0.879 & 0.864(2) & 0.879 & 0.878(1) \\
F & $(0.4662,0.936)$ & 0.6563 & 0.947 & 0.938(1) & 0.947 & 0.946(1) \\
\bottomrule
\end{tabular}
\end{table}

\begin{figure}
\centering
\includegraphics[width=\textwidth]{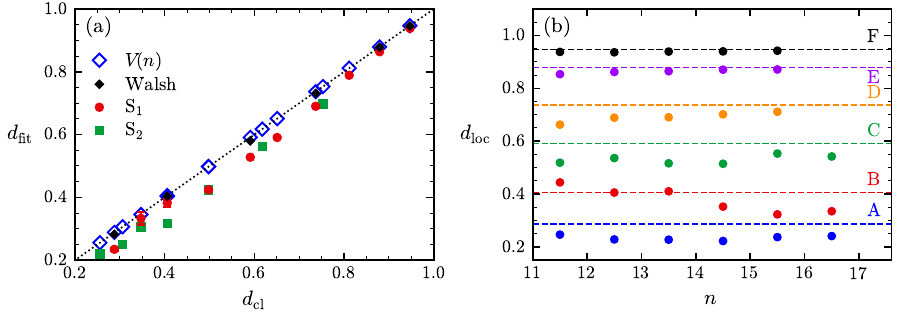}
\caption{Fractal Weyl law above $\wc$. (a) Fitted exponent $d_{\rm fit}$ versus $d_{\rm cl}=\htop/\ln2$,
equation~\eqref{eq:weyl}: Fourier baker at $c=0.2$ for the six windows of table~\ref{tab:weyl} and four further
one-strip windows, one of them on the route at $d_{\rm cl}=0.347$ (circles, S$_1$) and for seven two-strip windows (squares, S$_2$); exponent of the Walsh vertex
count $V(n)$ for all of them (open diamonds) and of the Walsh eigenvalue count at $c=0.2$ for A--F (small filled
diamonds, `Walsh'). The
jackknife errors of $d_{\rm fit}$ are $\le0.014$ except $0.020$ at $d_{\rm cl}=0.347$, so most error bars are hidden by
the symbols. Dotted line: $d_{\rm fit}=d_{\rm cl}$. (b) Local slope
$d_{\rm loc}=\log_2[\mathcal N_{2N}(0.2)/\mathcal N_N(0.2)]$, with $\mathcal N_N(c)$ the count of
equation~\eqref{eq:weyl} at size $N$: the slope of $\ln\mathcal N(0.2)$ versus $\ln N$ between the consecutive
dyadic sizes $N=2^n$ and $2N$, plotted at $n+\tfrac12$, for A--F, and $d_{\rm cl}$ (dashed, same colours). A law
$\mathcal N(c)\propto N^{d}$ gives $d_{\rm loc}=d$ at every $n$.}
\label{fig:weyl}
\end{figure}

In the Walsh quantisation \cite{NZ2005,NZ2007,KNNS2008}, a member of the family of generalised baker maps of
\cite{Ermann2006family}, the open map is a weighted adjacency matrix of the de Bruijn graph restricted to the window,
so its nonzero eigenvalues live on the non-trivial strongly connected components. The squared moduli of its
elements give the Ulam matrix of the open doubling map on $N$ cells, and Ulam approximations of the
Perron--Frobenius operators of open chaotic maps obey a fractal Weyl law of their own \cite{ErmannShepelyansky2010}.
The number $V(n)$ of vertices of these components, an upper bound on the number of nonzero eigenvalues, tests the
classical entropy: at $n=13$--$22$ its exponent equals $d_{\rm cl}$ to $3\times10^{-3}$ or better. The quantum test
is the eigenvalue count. The number of nonzero Walsh eigenvalues is $0.5$--$0.9$ of $V(n)$, because some eigenvalues
vanish in Jordan blocks, but the ratio hardly changes with $n$, and the exponent of the count lies within $0.011$ of
$d_{\rm cl}$ for windows A--F (table~\ref{tab:weyl}). Without diffraction, then, the count grows with the classical
exponent, as the Walsh spectrum follows the classical route all along.

The Fourier count grows more slowly. For the six windows its exponent lies below $d_{\rm cl}$ by $0.01$ (window F) to
$0.06$ (C), by more than two standard errors in five of them (table~\ref{tab:weyl} and figure~\ref{fig:weyl}). For B
the difference is smaller, but its local slope drifts from $0.44$ to $0.32$--$0.34$ at $N=2^{15}$--$2^{17}$
(figure~\ref{fig:weyl}(b)). Dyadic and non-dyadic $N$ give the same exponent within $0.02$ up to $N=2^{14}$, so the
deficit is not an arithmetic effect of $N=2^n$. Three further windows inside entropy plateaus, followed like the
window on the route, give deficits of $0.02$ to $0.07$ (figure~\ref{fig:weyl}(a)).

The deficit survives a change of cutoff: at $c=0.4$, $d_{\rm fit}$ agrees with its value at $c=0.2$ to
$0.035$ in all ten one-strip windows of figure~\ref{fig:weyl}(a) and in the seven two-strip windows below, and the
deficit remains in sixteen of these seventeen. In particular it does not come from eigenvalues
near $|\lambda|\approx0.36$, where the zero-entropy windows carry a group of diffractive resonances whose number grows
as $\log_2N$ \cite{ErmannP18}. A slowly varying prefactor,
$\mathcal N(c)\propto N^{d_{\rm cl}}(\ln N)^{-\omega}$, would make all the local slopes rise as $1/n$. C rises by
about the predicted amount, but A stays flat, B falls and D and E rise several times faster
(figure~\ref{fig:weyl}(b)), so a single logarithmic prefactor does not account for the six windows. Rigorous counts
exist for the class of maps of Dyatlov and Jin (section~\ref{sec:pressure}), for which they proved
$\#\{|\lambda|\ge b^{-\tau}\}=O(N^{\min(2\tau+2d-1,d)+\epsilon})$ \cite{DyatlovJin2017} and Li proved
$O(N^{d})$ \cite{Li2022}; these bounds do not apply to our maps.

Windows split into two strips behave in the same way. Two strips can trap a set of positive entropy even when their
total width is below $1/2$, where a single strip traps none (the lowest threshold over the strip positions is
$w=1/3$, for strips at distance $1/3$). For seven two-strip windows with $\htop=0.18$--$0.52$, three of
them with $w<1/2$, the Fourier exponents lie
$0.04$--$0.09$ below $d_{\rm cl}$, and the Walsh vertex count again gives $d_{\rm cl}$ to $2\times10^{-3}$
(figure~\ref{fig:weyl}(a)).

Like the precursor of section~\ref{sec:hybrid} and the excess of $|\lambda_0|$ below the threshold, the deficit is a
difference between the two quantisations: the Walsh count grows with the classical exponent, while the Fourier
exponent stays a few hundredths below it at the sizes we reach. A deficit of the same size, decreasing slowly with
$N$, appears for the Fourier count of a related window of entropy $\ln\varphi$ in \cite{ErmannP18}. Whether
diffraction leaves a permanent deficit or only a slow approach is open.

\setcounter{totalnumber}{1}
\section{Discussion}\label{sec:discussion}

We have studied the resonances of the open quantum baker map with a symmetric window, below and above the
classical transition to positive entropy. The resonance spectrum follows the classical route to chaos step by step,
also where the entropy is zero. The Thue--Morse cascade appears as a sequence of coexisting multiplets, each fixed by
an action phase, of which we resolve four. Near the threshold the cascade leaves a quantum precursor: one parity
sector of the 8-multiplet is perturbed by the 16-cycle through a return path that classically opens only at $\wc$,
with an amplitude of order $1/\Delta$, and the perturbation disappears when the sites of this path are removed from
the window. The transition itself appears as a jump of $|\lambda_0|$ that sets in within one or two lattice sites.
At $q_c=1/2$ and $0.35$, one site or more above the threshold, $|\lambda_0|$ lies between the classical value $\ee^{-\gamma/2}$ and the pressure value
$\ee^{P(1/2)}$, on average over $n=17$--$25$ within $2\%$ of the latter, with no free parameter. The jump is smaller
than the pressure would give, because $|\lambda_0|$ stays slightly below $\ee^{P(1/2)}$ above the threshold and
already exceeds $2^{-1/2}$ a few sites below it. Its height decreases slowly up to $N=2^{25}$. Above the threshold the Walsh eigenvalue count grows with the fractal
Weyl exponent, and the Fourier exponent lies $0.01$--$0.06$ below it in the six windows followed to
$N=2^{16}$--$2^{17}$ and up to $0.09$ below it in those followed to $2^{14}$.

The doubly exponential accumulation sets a quantum resolution limit on this route to chaos. A cycle is resolved when
its period is shorter than about $\log_2N$, twice the Ehrenfest time, so a Hilbert space of dimension $N$ resolves
$\nu_{\max}\simeq\log_2\log_2N$ steps of the cascade: four at $N=2^{18}$--$2^{25}$, while the fifth would need
$N\approx3\times10^{10}$. For a cascade that accumulates geometrically, as Feigenbaum's does, the same criterion
would give $\nu_{\max}\propto\log N$, and on a period-tripling route, whose thresholds also accumulate doubly
exponentially, it gives $\nu_{\max}\simeq\log_3\log_2N$ (appendix~\ref{app:routes}). Above $\wc$ the staircase limits
the resolution in the same way: the plateaus $\ln\varphi/2^j$ with $j\ge3$ and $\ln\theta/2^j$ with $j\ge2$ lie
within $7.2\times10^{-7}$ of $\wc$, less than one lattice site for $N\le2^{20}$ and $24$ sites at $N=2^{25}$. The
quantum map therefore resolves the transition only to about one lattice site, $1/N$, and sees it as a jump whose
classical part is the entropy of a window one site above $\wc$, bounded on both sides by
equation~\eqref{eq:hstair}. If $|\lambda_0|$
keeps following it and the excess below the threshold vanishes, the jump vanishes as $N\to\infty$, but only as
$1/\log_2N$. Our data up to $N=2^{25}$ do not distinguish this from a weak
power law, and from one $n$ to the next the quantum jump moves between $J_{\rm cl}$ and $J_P$, according to how close
$|\lambda_0|$ comes to the pressure value above the threshold and to $2^{-1/2}$ below it.

The symmetric window is not the only family of windows that ends at the Thue--Morse point. On every straight line
through that point along which the windows are nested, the cascade is the same, and only the distances at which the
cycles enter change, by a factor between $1$ and $2$ (appendix~\ref{app:routes}). What singles out $q_c=1/2$ is the
quantum problem. For even $M$ the lattice window is symmetric about $1/2$, so its ends lie on the line for every $N$.
On other lines one end of the window is fixed at an irrational point and the lattice puts it on one side or the other:
on the line $q_c-w/2=1-2a^*$, for instance, $|\lambda_0|$ is $0.750$--$0.767$ or $0.719$--$0.731$ depending on the side
on which the first site of the window falls ($N=2^{12}$--$2^{18}$). 

Several questions remain open. We have not derived the $1/\Delta$ law of the precursor. We do not know whether
$|\lambda_0|$ keeps following the entropy beyond $N=2^{25}$, which the $1/\log_2N$ limit of the jump assumes, nor
whether the excess of $|\lambda_0|$ over $2^{-1/2}$ a few sites below the threshold, which does not decrease up to
$N=2^{25}$ and also enters the jump, vanishes as $N\to\infty$. Finally, we do not know whether the deficit of the Fourier Weyl exponent closes:
at $N\le2^{17}$ the local slopes rise slowly towards $d_{\rm cl}$ in three of the six windows, stay roughly constant in
two and fall in one.

Natural extensions are asymmetric bakers and maps with more than two branches, where
the modulus of the multiplet depends on the cycle, and where we expect the cascade to map onto a devil's staircase in
the escape rate. Implementations of the open map on quantum circuits were proposed for single-orbit multiplets, and tested in pilot
runs on a superconducting processor, in \cite{Ermann2026PRL}.

\appendix

\section{Reduction of holes that contain the fixed point}\label{app:lemma}

The results of \cite{GlendinningSidorov2015} are stated for holes $(a,b)$ with $0<a<b<1$. Our window centred at
$q_c=1/2$ corresponds to a hole that contains the fixed point $0$, and the following elementary lemma reduces it to that
form.

\begin{lemma}\label{lem:reduction}
Let $u,v>0$ with $u+v\le1/2$, $H=(1-u,1)\cup[0,v)$ and $H'=\big((1-u)/2,\,(1+v)/2\big)$, so that $\sigma(H')=H$. Then
$\mathcal K(H)\subset\mathcal K(H')$ and $\mathcal K(H')\setminus\mathcal K(H)\subset\{0\}\cup\bigcup_{k\ge1}\sigma^{-k}\mathcal K(H)$. In particular, $\mathcal K(H)$ and
$\mathcal K(H')$ have the same Hausdorff dimension, and the restrictions of $\sigma$ to them have the same topological entropy.
\end{lemma}

\begin{proof}
If $\sigma^tx\in H'$ for some $t$, then $\sigma^{t+1}x\in H$; hence $\mathcal K(H)\subset\mathcal K(H')$. Next,
$\sigma^{-1}(H)=H'\cup[0,v/2)\cup(1-u/2,1)$, and the last two intervals lie in $H$. So if $x\in\mathcal K(H')$ and $\sigma^tx\in H$
for some $t\ge1$, then $\sigma^{t-1}x\in H$: the times at which the orbit of a point of $\mathcal K(H')$ is in $H$ form an initial
segment $\{0,\dots,t_0\}$. If $t_0$ is finite, $\sigma^{t_0+1}x\in\mathcal K(H)$. If it is infinite, the orbit never leaves $H$,
and only $x=0$ does so: a point of $(0,v)$ is doubled until it falls in $[v,2v)$, and a point of $(1-u,1)$ until it
falls in $(1-2u,1-u]$, both disjoint from $H$ because $u+v\le1/2$. The set $\sigma^{-k}\mathcal K(H)$ is the union of $2^k$ images
of $\mathcal K(H)$ under similarities of ratio $2^{-k}$, which gives the equality of dimensions. Finally, the points of
$\mathcal K(H')\setminus\mathcal K(H)$ lie in $H$, and those other than $0$ visit $H$ only finitely often, so by Poincar\'e recurrence
every invariant probability measure on $\mathcal K(H')$ gives them zero weight. The fixed point carries zero entropy, and the
variational principle gives the equality of entropies.
\end{proof}

\section{Origin and exact certificates of the entropy plateaus}\label{app:cert}

The plateaus of section~\ref{sec:entropy} come from the following results of \cite{Nilsson2010}. Theorems~66, 63
and~26 give the continuity of the dimension, its constancy on intervals of full measure and its vanishing for
$h\ge1-2a^*$. Lemma~61 and Theorem~64 show that the interval attached to a minimal word is a maximal interval of
constancy, Theorem~57 is the Thue--Morse substitution that divides the dimension by a power of two from one range of
$h$ to the next, and Lemma~8 gives $\dim_HF=\log_2\varphi$ for the word $001$. In terms of $h=(1-w)/2$,
equation~\eqref{eq:plateau} is the interval $[341/2016,\,11/63]$ of the word $001011$,
equation~\eqref{eq:plateau2} is $[489233/2795520,\,239/1365]$ (the word $001011001101$), the plateau $\ln\theta$ of
equation~\eqref{eq:plateau3} is the interval of the word $00101$, and the plateaus $\tfrac12\ln\theta$ and
$\tfrac14\ln\theta$, on $665/1023\le w\le170419/261888$ and $227127/349525\le w\le0.6498168470$, are those of the
words $0010110011$ and $00101100110100101101$. In the classification of
\cite{AlcarazBakerKong}, $X_h$ is their $V_q$, and their Theorem~2 gives $w=21/31$ and $173/248$ from the irreducible
word $11010$, and the edges $41/63$ and $667/1008$ of equation~\eqref{eq:plateau} from the $*$-irreducible word
$110100$.

We checked the endpoints in equations~\eqref{eq:plateau} and~\eqref{eq:plateau2}, which we obtained by evaluating
Nilsson's chain, independently. A bracket with guaranteed bounds is obtained from the inner and outer SFTs of section~\ref{sec:boundary}, written
as functions of $h$ for the surviving interval $[h,1-h]$. Since $X_h$ decreases with $h$,
\begin{equation}
\htop\!\left(\mathrm{SFT}^{\rm in}_m(h_2)\right)\le\htop(h)\le\htop\!\left(\mathrm{SFT}^{\rm out}_m(h_1)\right),
\qquad h_1\le h\le h_2 .
\end{equation}
The spectral radius of the two adjacency matrices is computed exactly. The graphs have up to $3.3\times10^4$
vertices, most of them carrying only the eigenvalue zero, so we avoid floating point altogether. First, vertices
without predecessors or successors are removed. Second, two vertices with equal rows (or equal columns) of the
adjacency matrix $A$ are merged: if $A=RS$, with $S$ the matrix of distinct rows and $R$ the $0/1$ matrix that assigns
each vertex to its class, then $SR$ has the same nonzero spectrum as $A$ (an elementary strong shift equivalence).
Iterating these two steps reduces each graph to a non-negative integer matrix of size $8$ to $106$. Its
characteristic polynomial is computed in rational arithmetic and divided exactly by $x^4-x^2-1$ (largest root
$\sqrt\varphi$) or by $x^8-x^4-1$ (largest root $\varphi^{1/4}$). The quotient is a product of $x^k$, cyclotomic
polynomials and, for some outer graphs, $x^{12}-x^6-1$ or $x^{18}-x^{12}-x^6-1$, whose roots have
$|x|\le2^{1/6}=1.122$ by Cauchy's bound in $y=x^6$, below both $\varphi^{1/4}=1.128$ and $\sqrt\varphi=1.272$. As a check, for each of the $34$ graphs of
the certificates of equations~\eqref{eq:plateau}, \eqref{eq:plateau2} and~\eqref{eq:plateau3} and of the plateaus
$\tfrac12\ln\theta$ and $\tfrac14\ln\theta$, $\mathrm{tr}\,A^k$ of the full and of the reduced matrix and the power sums of the roots of the
characteristic polynomial agree for $k\le60$, all in integer arithmetic, and for $26$ of them a direct count of the
periodic words of period $k\le24$ whose $m$-blocks are allowed gives $\mathrm{tr}\,A^k$ again. For equation~\eqref{eq:plateau} we applied this at $m=16,18,\dots,24$ with $h_1,h_2$
at the inner edges of the plateau found by bisection, and for both plateaus at $m=20,24$ with $h_1,h_2$ at distance
$2^{2-m}$ inside the endpoints above; all graphs pass. Bisection at $m=24$ brackets the edges of
equation~\eqref{eq:plateau} to $1.2\times10^{-7}$, and the brackets contain $41/63$ and $667/1008$; just outside the
intervals the entropy is different. The lower edge $41/63=1-2\cdot11/63$ is the appearance threshold of the
period-6 cycle $001011$. So our numerical data reproduce Nilsson's classification on the plateaus that matter
for the quantum problem.

For the plateau $\ln\theta$, at
$m=14,16,18,20,24$ the characteristic polynomials of $\mathrm{SFT}^{\rm in}_m$ at $h=5/31-2^{2-m}$ and of
$\mathrm{SFT}^{\rm out}_m$ at $h=75/496+2^{2-m}$ are divisible by $x^4-x^3-x^2+x-1$, and the quotient is a product of
cyclotomic polynomials or has roots of modulus at most $1.13<\theta$. The same test, with $x^4-x^3-x^2+x-1$
evaluated at $x^2$ and at $x^4$, certifies the plateaus $\tfrac12\ln\theta$ ($m=20,24$) and $\tfrac14\ln\theta$
($m=26$) at distance about $2^{2-m}$ inside their edges; there the quotient is a product of powers of $x$ and
cyclotomic polynomials, so that no other root competes with $\theta^{1/2}=1.230$ or $\theta^{1/4}=1.109$.

Away from $q_c=1/2$ the plateaus are regions of the plane, bounded by lines of slope $\pm2$, on which one end of the
hole is fixed (figure~\ref{fig:cl}(a)). On the grid of $2001\times2001$ windows the fourteen largest plateaus cover
$29.4\%$ of the map, and on each of them $e^{\htop}$ agrees to $10^{-9}$ with the largest root of a polynomial with
coefficients $0$ and $\pm1$, such as $x^3-x^2-1$ ($6.4\%$ of the map) and $x^2-x-1$ ($6.3\%$); these identifications
are numerical. The plateaus $\tfrac12\ln\varphi$ and $\ln\theta$ of figure~\ref{fig:cl}(b) cover only $0.01\%$ and
$0.13\%$.

\section{The operator of \texorpdfstring{$\lambda_{\rm D}$}{lambda\_D}}\label{app:D}

Let $\mu$ be a signed half-integer and $p_\mu=\mu/N$ the momentum measured in lattice units from the discontinuity
$p=0$ of the map, with $\mu<0$ just below $p=1\equiv0$. For a window that does not contain $q=0$ the operator of
section~\ref{sec:mult} is
\begin{equation}
\mathcal D_{\mu\mu'}=\sqrt2\,\mathrm{sgn}(\mu')\int_{I_+}\ee^{2\pi\mathrm i(\mu'-2\mu)x}dx\ \ (\mu>0),\qquad
\mathcal D_{\mu\mu'}=-\sqrt2\,\ee^{\mathrm i\pi|\mu'|}\int_{I_-}\ee^{2\pi\mathrm i(\mu'-2\mu)x}dx\ \ (\mu<0),
\label{eq:D}
\end{equation}
with $I_+=W\cap[0,\frac12)$ and $I_-=(W\cap[\frac12,1))-\frac12$. It comes from the matrix elements of $U\Pi_W$
between the momentum states $|p_\mu\rangle$, which are exactly
$\sqrt2\,\mathrm{sgn}(\mu')N^{-1}\sum_x\ee^{2\pi\mathrm i(\mu'-2\mu)x}$ for $\mu>0$ and
$-\sqrt2\,\ee^{\mathrm i\pi|\mu'|}N^{-1}\sum_x\ee^{2\pi\mathrm i(\mu'-2\mu)x}$ for $\mu<0$, with $x$ running over the site
centres $(j+\frac12)/N$ in $I_+$ and $I_-$ (we checked this identity numerically to $10^{-13}$). These are midpoint
sums of the integrals in equation~\eqref{eq:D}, from which they differ by at most
$\sqrt2\pi^2(\mu'-2\mu)^2/(12N^2)$, so at fixed $\mu,\mu'$ the matrix converges entrywise to $\mathcal D$. The rows with
$\mu>0$ integrate over the left part of the window and those with $\mu<0$ over the right part, because the left half of
the square sends momenta near $p=0$ to momenta just above $p=0$, and the right half sends momenta near $p=1$ to
momenta just below $p=1$, in both cases halving the offset; this is why the integrand peaks at $\mu'=2\mu$. Since the
window cuts the integrals sharply, every $\mu'$ is coupled to every $\mu$, with amplitudes that decay only as
$1/|\mu'-2\mu|$. Truncated to $|\mu|<512$, the leading eigenvalue of $\mathcal D$ is converged to $10^{-5}$.

Entrywise convergence does not imply that the spectrum of a non-normal operator converges, so the identification of
the block eigenvalue $\lambda_{\rm D}$ with the leading eigenvalue of $\mathcal D$ is numerical. In the windows
near $\wc$ of section~\ref{sec:mult} the distance between the two does not decrease steadily with $N$: it is below
$10^{-4}$ at $n=23$, in the windows that end $22$ to $42$ sites per side below $N\wc$, and $2.3\times10^{-3}$ and
$8\times10^{-4}$ at $n=24$ and $25$; within three sites per side of the threshold it scatters between
$2\times10^{-3}$ and $0.1$. Besides these windows
near $\wc$ of section~\ref{sec:mult}, we followed it to larger $N$ in two windows of width $1/2$ (Arnoldi up to
$N=2^{24}$). At $q_c=1/2$, where the window traps the 2-cycle at the start of the route, $\mathcal D$ gives
$0.28025-0.40442\,\mathrm i$, and the distance to the block eigenvalue decreases from $7\times10^{-2}$ at $n=8$ to
$3.1\times10^{-4}$ at $n=24$, with an alternation between even and odd $n$. For the asymmetric window at $q_c=0.356$,
which traps the 3-cycle $001$, $\mathcal D$ gives $0.37355-0.42092\,\mathrm i$, and the distance stays at
$1.2\times10^{-3}$ from $n=19$ on, so there $\mathcal D$ describes $\lambda_{\rm D}$ only to this accuracy.

\section{Other routes and other lines to the boundary}\label{app:routes}

In the plane of the window ends, $q_\pm=q_c\pm w/2$, other routes reach the boundary of the zero-entropy region
through cascades in which the period is multiplied by the denominators $m_j$ of a sequence of rationals
(section~4.3 of \cite{GlendinningSidorov2015}), and by Lemma~3.2 of \cite{GlendinningSidorov2015} the thresholds then
approach the boundary as $2^{-Q_j}$, with $Q_j=m_1\cdots m_j$. We checked one period-tripling route, the sequence
$(1/3,1/3,\dots)$. On the line $q_-=q_-^\infty=0.071550880\ldots$ the only cycles of period $L\le30$ that survive below
$q_+^\infty=0.634905376\ldots$ have periods $3$, $9$ and $27$. The cycle of period $3^j$ enters at $q_+^{(j)}$, with
$q_+^\infty-q_+^{(j)}=0.44\times2^{-3^j}$ for $j\ge2$, and already at $q_+=q_+^\infty+10^{-4}$ the entropy is
$\htop\ge0.11$. The resolution criterion of section~\ref{sec:mult}, a cycle two or three sites inside the window, now
applies to the single moving edge, $N(q_+^\infty-q_+^{(j)})\gtrsim2.5$, and gives $\nu_{\max}\simeq\log_3\log_2N$: the
9-cycle needs $N\gtrsim3\times10^3$ and the 27-cycle $N\gtrsim8\times10^8$.

Along a straight line in the plane $(q_-,q_+)$ on which $q_-$ does not increase and $q_+$ does not decrease, the
windows are nested, so the survivor set grows, $\htop$ is monotone and the line crosses the boundary at a single point
$B$. For a window that does not contain $0$, a cycle survives if and only if $q_-\le\min\mathcal O$ and $q_+>\max\mathcal O$, where
$\mathcal O$ is the set of its points. The cycles that survive just below the transition therefore depend only on $B$, and the
slope of the line only changes the distances at which they enter. At the Thue--Morse point, $B=(1-2a^*,2a^*)$, both
$B$ and the Thue--Morse cycles are symmetric under $x\to1-x$, so $\min\mathcal O-(1-2a^*)=2a^*-\max\mathcal O\equiv d_{\mathcal O}$ for every
cycle. On a line through $B$ with slope $dq_+/dq_-=-\eta$ the cycle then enters at $\wc-w=d_{\mathcal O}\,[1+\min(\eta,1/\eta)]$:
the cascade is the same on every such line, up to a factor between $1$ and $2$. We checked this for the cycles of
period $2$, $4$, $8$ and $16$ on the line $q_-=1-2a^*$ ($\eta=\infty$), where $\wc-w$ is exactly half of its value
at $q_c=1/2$ ($\eta=1$), from $0.158$ for the 2-cycle to $5.3\times10^{-6}$ for the 16-cycle. 

A line that misses $B$ crosses the boundary at another point, in general after a finite chain of renormalisations,
and its cascade is truncated. On the line $q_c=0.4975$, for instance, only the cycles $01$ and $0011$ survive below the
threshold (all periods up to $18$), and the threshold given by \cite{GlendinningSidorov2015} equals $2(q_c-7/40)$ to
$10^{-13}$. This law was found for repellers made of a few orbits in \cite{ErmannCarlo2026}, where these windows are
studied; we checked it at four other values of $q_c$ between $0.492$ and $0.4992$.

\ack{Claude (Anthropic) assisted with numerical code, figures and text, and Gemini (Google) with critical reading;
the author directed the work, verified the results, and is solely responsible for the content.}

\funding{This work was supported by CONICET under Grant No.\ PIP 2022-2024 GI -- 11220210100208CO.}

\data{The certificate scripts of appendix~\ref{app:cert} are ancillary files of the arXiv version; other data and
codes are available from the author upon reasonable request.}

\end{document}